\documentclass[12pt, draftclsnofoot, onecolumn]{IEEEtran}
\usepackage{graphicx,amsmath,amssymb}
\usepackage{subfigure}
\usepackage{citesort}
\usepackage{fancyhdr}
\usepackage{mdwmath}
\usepackage{mdwtab}
\usepackage{balance}
\usepackage{xcolor}
\usepackage{bm}
\usepackage{amsthm}
\usepackage{algorithm}
\usepackage{multirow}
\usepackage{flafter}
\usepackage{setspace}
\usepackage{cite}
\usepackage{algpseudocode}

\newtheorem{definition}{Definition}
\newtheorem{remark}{Remark}

\newtheorem{theorem}{Theorem}

\newtheorem{lemma}{Lemma}

\newtheorem{corollary}{Corollary}

\newtheorem{assumption}{Assumption}

\begin{document}
\title{Resource Allocation in Uplink NOMA-IoT Networks: A Reinforcement-Learning Approach}

\author{Waleed~Ahsan,~\IEEEmembership{Student Member,~IEEE,}
        Wenqiang~Yi,~\IEEEmembership{Member,~IEEE,}
        Zhijin~Qin,~\IEEEmembership{Member,~IEEE,}
        Yuanwei~Liu,~\IEEEmembership{Senior Member,~IEEE,}
        and Arumugam~Nallanathan,~\IEEEmembership{Fellow,~IEEE}
\thanks{W. Ahsan, W. Yi, Z. Qin, Y. Liu, and A. Nallanathan are with Queen Mary University of London, London, UK (email:\{w.ahsan, w.yi, z.qin, yuanwei.liu, a.nallanathan\}@qmul.ac.uk).
\par Part of this work was submitted in IEEE International Conference on Communications (ICC) Workshops, June, Ireland, 2020~\cite{123456789}.
\par This work was supported by the U.K. Engineering and Physical Science Research Council (EPSRC) under Grant EP/R006466/1. 
}
}

\maketitle
\vspace{-2 cm}
\begin{abstract}
  Non-orthogonal multiple access (NOMA) exploits the potential of the power domain to enhance the connectivity for the Internet of Things (IoT). Due to time-varying communication channels, dynamic user clustering is a promising method to increase the throughput of NOMA-IoT networks. This paper develops an intelligent resource allocation scheme for uplink NOMA-IoT communications. To maximise the average performance of sum rates, this work designs an efficient optimization approach based on two reinforcement learning algorithms, namely deep reinforcement learning (DRL) and SARSA-learning. For light traffic, SARSA-learning is used to explore the safest resource allocation policy with low cost. For heavy traffic, DRL is used to handle traffic-introduced huge variables. With the aid of the considered approach, this work addresses two main problems of fair resource allocation in NOMA techniques: 1) allocating users dynamically and 2) balancing resource blocks and network traffic. We analytically demonstrate that the rate of convergence is inversely proportional to network sizes. Numerical results show that: 1) Compared with the optimal benchmark scheme, the proposed DRL and SARSA-learning algorithms have lower complexity with acceptable accuracy and 2) NOMA-enabled IoT networks outperform the conventional orthogonal multiple access based IoT networks in terms of system throughput.
  \end{abstract}
\begin{IEEEkeywords}
{D}eep reinforcement learning,  internet of things,  non-orthogonal multiple access, power allocation, SARSA learning, user clustering
\end{IEEEkeywords}
\section{Introduction}
\vspace{-0.2 cm}
  Internet of things (IoT) enable millions of devices to communicate simultaneously. It is predicted that the number of IoT devices will rapidly increase in the next decades~\cite{zhai2019delay}. Owing to a large number of time-varying communication channels, the dynamic network access with massive connectivity becomes a key requirement for future IoT networks. Recently, non-orthogonal multiple access (NOMA) is evolved as a promising approach to solve this problem\cite{islam2017power},\cite{sharma2019towards}. The key benefit of using NOMA is that NOMA exploits the power domain to enable more connectivity than the traditional orthogonal multiple access (OMA).  More specifically, NOMA supports multiple users at the same time/frequency resource block (RB) by employing superposition coding at transmitters and successive interference cancellation (SIC) techniques at receivers \cite{wan2018non}. Various model-based schemes have been proposed to improve different metrics of NOMA-IoT networks, such as coverage performance, energy efficiency, system throughput (sum-rates), etc. Additionally, on the importance of sum-rates, the recent work in wireless networks based on the state of the art reflective intelligent surfaces (RIS) considered sum-rate maximization objective function \cite{guo2020weighted}. The sum-rate is an important parameter to depict the average performance of wireless networks in detail for each user. Due to this, the sum-rate is widely used as a significant performance indicator for wireless networks by numerous research works \cite{zeng2020sum}, \cite{tse2005fundamentals}. It shows the significance of the sum-rate maximization based objective functions. Regrading the system design, the uncertainty and dynamic mechanisms of wireless communication environments are difficult to be depicted by an accurate model. The dynamic mechanism involves spectral availability, channel access methods (e.g., OMA, NOMA, hybrid systems, etc.), and dynamic traffic arrival. Especially in practical NOMA systems by allowing resource share among more than one users the process is more dynamic when users are simultaneously joining and leaving the network in short term and long term basis. Numerous model-based techniques target to solve dynamic behaviour of wireless networks but failed to provide long-term performance outcomes \cite{ding2017survey}, \cite{shao2018dynamic}, \cite{ali2016dynamic},\cite{miuccio2020joint} and \cite{mostafa2019connection}. Moreover, due to the absence of learning abilities, to provide long term network stability the computational complexity of traditional schemes becomes ultra-high. This is due to the fact that, by default, traditional approaches cannot extract knowledge from any given problem (e.g, given distributions) online. Fortunately, the online learning properties of recently developed machine learning (ML) methods are extremely suitable to handle such type of dynamic problems \cite{8519960}.
\subsection{Related Works and Motivations}
\subsubsection{Studies on NOMA-IoT Networks}
Due to the aforementioned benefits, academia has proposed numerous studies on the optimization of resource allocation in NOMA-enabled IoT networks. For single-cell scenarios, the authors in \cite{shao2018dynamic} proposed a two-stage NOMA-based model to optimize the computation offloading mechanism for IoT networks \cite{hussain2019machine}. In the first stage, a large number of IoT devices are clustered into several NOMA groups depending on their channel conditions. In the second stage, different power levels are allocated to users to enhance network performance. The comparison between uplink NOMA-IoT and OMA-IoT is presented in \cite{zhang2016uplink}, which considered the optimal selection of targeted data rates for each user. Regarding downlink transmission, the similar topic was studied in~\cite{ding2014performance} and~\cite{hanif2016minorization}.  Different from others, in \cite{zhang2018energy} using 2D matching theory authors performed dynamic resource allocations considering energy efficiency for downlink NOMA. Similarly, in \cite{miuccio2020joint} for the massive Machine Type Communications (mMTC) usage scenario, also known as massive Internet of Things (mIoT) dynamic resource management is performed with Sparse Code Multiple Access (SCMA) domain using conventional mathematical tools. The authors in \cite{yang2016general} proposed a general power allocation scheme for uplink and downlink NOMA to guarantee the quality of service (QoS). In \cite{zhai2018energy}, NOMA scheduling schemes in terms of power allocation and resource management were optimized to realize the massive connectivity in IoT networks. For multi-cell scenarios, the impact of NOMA on large scale multi-cell IoT networks was investigated in \cite{liu2017enhancing}. To characterize the communication distances, the authors in~\cite{8635489} analysed the performance of large scale NOMA communications via stochastic geometry. It is worth noting that NOMA-IoT channels are time-varying in the real world. Therefore, the study in \cite{ali2018coordinated} considered a practical framework with dynamic channel state information for evaluating the performance of massive connectivity. The authors in \cite{qian2018optimal},\cite{shahab2019grant}, and \cite{dai2018survey} discussed the advantages of various NOMA-IoT applications. Interestingly, the proposed schemes introduced artificial intelligence (AI) methods to solve some practical challenges of NOMA-IoT systems. For both uplink and downlink scenarios, AI-based multi-constrained functions can be utilized to optimise multiple parameters simultaneously.
\subsubsection{Studies on ML-based NOMA Systems}
Due to the dynamic nature of NOMA-IoT communications, traditional methods may not be suitable for such type of networks \cite{mostafa2019connection}. Note that ML-based methods are capable to handle the complex requirement of future wireless networks via learning. In \cite{gui2018deep}, one typical deep learning method, namely long short-term memory (LSTM) \cite{hochreiter1997long}, was applied for the maximization of user rates by minimizing the received signal-to-noise-ratio (SINR). In \cite{xu2018outage}, a successive approximation based algorithm was proposed to minimize outage probabilities through optimizing power allocation strategies. For next-generation ultra-dense networks, ML-aided user clustering schemes were discussed in \cite{jiang2017machine} for obtaining efficient network management and performance gains. Because using clustering schemes, the entire network can be divided into several small groups, which helps to ease resource management~\cite{bi2015wireless}. Regarding AI-based cluster techniques, in \cite{arafat2019localization} and \cite{cui2018unsupervised}, resources were assigned to the most suitable user to ensure the best QoS for unmanned aerial vehicle (UAV) networks and millimetre wave networks, respectively. It is worth noting that the optimization of clustering is an NP-hard problem. Therefore, for such type of problems the authors in \cite{gui2018deep}, \cite{jiang2017machine}, and \cite{liu2019machine} recommended to use AI instead of conventional mathematical models. Currently, realistic datasets are not available for most of the machine learning algorithms, to overcome these designers use the synthetic dataset for simulations. The data set is generated for a certain environment so it is difficult to depict general property and online scenarios of wireless networks. Therefore, algorithms like reinforcement learning play a very important role where data is collected online (during simulation) to learn the given search space for the simulation requirements. There are various Q-learning algorithm variants used for NOMA systems. Due to inefficient learning mechanism, other methods like traditional Q-learning and Multi-arm bandits (MABs) are heavily influenced by regret (negative reward) \cite{li2020multi}\cite{de2018comparing}. On the other hand two most powerful methods, deep reinforcement learning (DRL) and SARSA learning created by google deep mind\cite{silver2017mastering} and by the authors in \cite{rummery1994line}. Both DRL deep mind and SARSA learning algorithms are efficient learners. Due to unique learning behaviour, DRL and SARSA tend to receive more rewards. The main advantage of the deep mind and online SARSA learning is to handle dynamic control as in \cite{lillicrap2015continuous}. With the development of such type of RL techniques, the challenges for NOMA systems, which are difficult to be solved via traditional optimization methods, have been reinvestigated via RL-based approaches~\cite{xiao2017reinforcement, liu2019uav, yang2019reinforcement}.
\subsubsection{Motivations}
Combining multi-user relationship and resource allocation increases the complexity of NOMA-IoT systems, which also introduces new problems for optimizing power allocation and scheduling schemes. Unlike traditional methods \cite{zhai2018energy}, where only one BS is considered for small scale network with no inter-cell interference and dynamic user connectivity. The design of schedulers should be in tandem with the large scale dynamic resource allocations and user decoding strategies. Therefore, due to the high complexity of the problem under multi-cell multi-user cases, AI can be a feasible option for the dynamic resource allocation~\cite{cui2017optimal}. For large-scale NOMA-IoT networks, an intelligent reinforcement learning (RL) algorithm becomes a promising approach to find the optimal long-term resource allocation strategy. This algorithm should jointly optimize multiple criteria under dynamic network states. In this paper, our main goal is to address the following research questions:
\begin{itemize}
  \item \textbf{Q1}: In NOMA-IoT networks, how to maximize the long-term sum rates of users for a given network traffic density?
  \item \textbf{Q2}: How does the inter-cell interference affect the long-term sum rates?
  \item \textbf{Q3}: What is the correlation between traffic density, system bandwidth, and the number of clusters in NOMA-IoT networks?
  \end{itemize}
 From above as it is known that model-free methods are suitable to address multi-constrained long-term problem online. Therefore, in long-term, there is a strong correlation of mentioned research questions with general problems of ``intermittent connectivity of IoT users (continuously joining and leaving the network), balanced resource allocations ( optimal allocations policy for dynamic network settings) and network traffic (as the (Min-Max) number of users competing for the resource blocks)" in wireless networks. Similarly, research Q1 for capacity maximization, research Q2 for network scalability and, research Q3 for long-term network performance is strongly dependent on the main problems ``balancing of network resources, IoT users and, the dynamic network behaviour". 
 \vspace{-0.2 cm} 
\subsection{Contributions and Organization}
This paper considers uplink NOMA-IoT networks, where multiple IoT users are allowed to share the same RB based on NOMA techniques. With the aid of RL methods, we propose a multi-constrained clustering solution to optimize the resource allocation among IoT users, base stations (BSs), and sub-channels, according to the received power levels of IoT users. Appropriate bandwidth selection for the entire system with different traffic densities is also taken into consideration for enhancing the generality. Our work provides several noteworthy contributions:
\begin{itemize}
\item We design a 3D association model-free framework for connecting IoT users, BSs, and sub-channels. Based on this framework, we formulate a sum-rate maximization problem with multiple constraints. These constraints consider long-term variables in the proposed NOMA-IoT networks, such as the number of users, channel gains, and transmit power levels. To characterize the dynamic nature (online), at each time slot, these variables are changeable.
\item We propose two RL techniques, namely SARSA-learning with $\epsilon-greedy$ and DRL, to solve this long-term optimization problem. SARSA-learning is used for light traffic scenarios to avoid high complexity and memory requirements. Heavy traffic scenarios with a huge number of variables are studied by DRL, where three different neuron activation mechanisms, namely TanH, Sigmoid, and ReLU, are compared to evaluate the impact of neuron activation on the convergence of the proposed DRL algorithm.
\item We design novel 3D state and action spaces to minimise the number of Q-tables for both SARSA and DRL frameworks. The considered action space represent switching between RBs, which is the most efficient strategy for our networks. Based on this adequate Q-table design, DRL is able to converge faster.
\item We show that: 1) according to the time-varying environment, resources can be assigned dynamically to IoT users based on our proposed framework; 2) for the proposed model, the learning rate $\alpha=0.75$ provides the best convergence and data rates; 3) for SARSA and DRL the sum-rate is proportional to the number of users; 4) DRL with the ReLU activation mechanism is more efficient than TanH and Sigmoid, and 5) IoT networks with NOMA provide better system throughput than those with OMA.
\end{itemize}
The rest of the paper is organised as follows: In Section II, the system model for the proposed NOMA-IoT networks is presented. In Section III, SARSA-learning and DRL-based resource allocation is investigated. The corresponding algorithms are also presented. Finally, numerical results and conclusions are drawn in Section IV and Section V, respectively.
\vspace{-0.3 cm}
\begin{figure*}[t!]
\centering
\includegraphics[scale=.4,keepaspectratio]{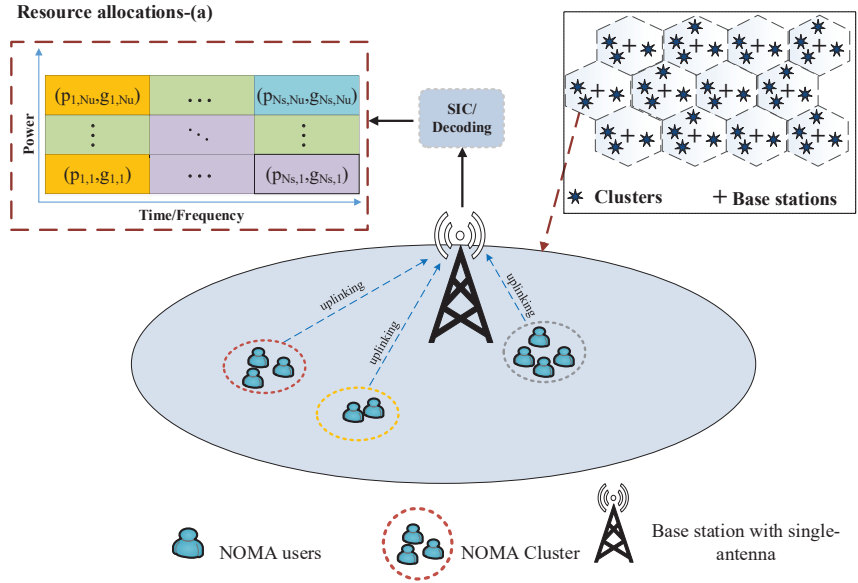}\\
\caption{Illustrating uplink NOMA resource allocation by using the optimization algorithm to efficiently cluster users for resource blocks at the base-station side. Resource allocations-(a) presents different resource blocks in yellow, green, and blue with power on (x-axis) and time/frequency on (y-axis) assigned to IoT users. The powers and gains of $N_u$ users are denoted with $p$ and $g$.}
\label {fig:intro}

\end{figure*}

\begin{table*}[htb!]
\tiny
\centering
\caption{Table of notations}
\label{tab11}
\begin{tabular}{l|l|l|l}
\hline
\hline
   \textbf{Symbol} &  \textbf{Definition} &   \textbf{Symbol} &  \textbf{Definition} \\ \hline
   \text{$N_b,b_i$} &  Number of BSs, symbol of BSs&\text{$N_s,s_j$} &  Number of sub-channels (NOMA clusters), symbol of sub- channels (NOMA clusters)   \\ \hline
   \text{$N_u,u_k$} &  Number of users, symbol of users& \text{$\Phi_u^{i,j}$,$u^{i,j}_k$} & Set of users connected to BS $b_i$ via sub-channel $s_j$, \ user $k$ in the set $\Phi_u^{i,j}$   \\ \hline	
  \text{$\Phi_b$} &  Set of BSs& \text{$c_k^{i,j}(t)$} & Clustering variable for user $u_k$ connecting to BS $b_i$ via sub-channel $s_j$ at time $t$  \\ \hline
   \text{$p_k^{i,j}(t)$} & Transmit power for user $u^{i,j}_k$ at time $t$&   \text{$g_k^{i,j}(t)$} & Channel gain for user $u^{i,j}_k$ at time $t$  \\ \hline
   \text{$\sigma(t)$} & Additive white Gaussian noise at time $t$&    \text{$I_{inter}(t)$} & Inter-cell interference at time $t$\\ \hline
   \text{$\gamma^{i,j}_{k}(t)$} & Instantaneous SINR for user $u^{i,j}_k$ at time $t$&   \text{$R^{i,j}_{k}(t)$} & Instantaneous data rate for user $u^{i,j}_k$ at time $t$\\ \hline
   \text{$R^{th}_{k}$} & Rate requirement for the SIC process of user $u^{i,j}_k$ & \text{$U_{s}$, $P_s$} & Maximal load of each sub-channel, Maximal power for each sub-channel \\ \hline
   \text{$T$} & Duration of the considered long-term communication&   \text{$\mathbf{C}$, $\mathbf{P}$} & Matrix for clustering parameters, matrix for transmit power\\ \hline
   \text{$\mathbf{ \theta_{t}}$} & Vector for DRL gradients&   \text{$\beta_1,\beta_2$} & Moment estimation decay rate\\ \hline

\hline
\end{tabular}
\end{table*}
\vspace{-0.2 cm}
\section{System Model}
In this paper, we consider an uplink IoT network with NOMA techniques as shown in Fig. \ref{fig:intro}, where $N_b$ BSs communicate with $N_u (t)$ IoT users via $N_s$ orthogonal sub-channels. we assume $N_u (t)$ dynamic in each time-slot in our model, however for simplicity we omit $(t)$ for further sections. Additionally, channel gains are also dynamic for each user at each time-slot, even for the same user. The BSs and sub-channels are indexed by sets $\Phi_b = \{b_1,...,b_{N_b}\}$ and $\Phi_s = \{s_1,...,s_{N_s}\}$, respectively. Regarding users, the set for users severed by one BS $b_i\in \Phi_b$ $(i\in[1,N_b])$ through a sub-channel $s_j\in\Phi_s$ $(j\in[1,N_s])$ is defined as $\Phi_u^{i,j}=\{u_1,...,u_{N_{u}^{i,j}}\}$, where $N_u^{i,j}$ is the number of the intra-set users and $\sum\limits_{i = 1}^{{N_b}} \sum\limits_{j = 1}^{{N_s}}N_u^{i,j}=N_u$. BSs and users are assumed to be equipped with a single antenna. For each BS, the entire bandwidth $B$ is equally divided into $N_s$ sub-channels and hence each sub-channel has $\frac{B}{N_s}$ bandwidth. In a time slot, we assume a part of users are active and the rest users keep silence. To share knowledge, we consider fiber link with ideal back-haul for inter BS connectivity. The defined notations in this system model are listed in TABLE~\ref{tab11}.
\subsection{NOMA Clusters}
 Based on the principles of NOMA, more than two users can be served in the same resource block (time/frequency), which forms a NOMA cluster. In this paper, each sub-channel represents one NOMA cluster and $N_u^{i,j} \ge 2$ \cite{kiani2018edge}. To simplify the analysis, we assume BSs contain perfect CSI of all users. That CSI is our state space showing signalling and the channel conditions of IoT users connected to sub-channel via base-station. A detailed explanation is present in section III-b and section III-c. Based on such CSI, BSs are capable to dynamically optimize the sub-channel allocation for active users in long-term communication. For an arbitrary user $u_k$, we define its clustering variable at time $t$ as follows:
\begin{align}\label{f}
c^{i,j}_{k}(t)=
\Big\{
\begin{tabular}{ccc}
  1, & user $u_k$ connects to BS $b_i$ via sub-channel $s_j$\\
  0, & otherwise
\end{tabular}.
\end{align}
It is worth noting that $c^{i,j}_k(t)$ also implies the activity status of users. If user $k$ is inactive, we obtain that $c^{i,j}_k(t)\equiv 0,\forall i, j$. The set of clustering parameters is defined as $\mathbf{C}_t$ and $c^{i,j}_{k}(t) \in \mathbf{C}_t, \forall i,j,k$.
\subsection{Signal Model}
In a NOMA cluster $s_j$, one BS $b_i$ first receives the superposed messages from the active users in $\Phi_u^{i,j}$ and then applies SIC to sequentially decode each user's signal~\cite{liu2016cooperative}.  Without loss of generality, we assume the order of channel gains is $g^{i,j}_1 \leq g^{i,j}_2,...,\leq g^{i,j}_{u_{N_{u}^{i,j}}}$, where $g^{i,j}_{k}$ is the channel gain for the $k$-th user in $\Phi_u^{i,j}$ \cite{803503}. Therefore, the decoding order in this paper is the reverse of the channel gain order~\cite{8680645}. In a time slot $t$, the instantaneous signal-to-interference-plus-noise ratio (SINR) for the intra-cluster user $u^{i,j}_k\in\Phi_u^{i,j}$ is given by
\begin{align}\label{a}
\gamma^{i,j}_{k}(t)= \frac{c^{i,j}_{k}(t)p^{i,j}_{k}(t) g^{i,j}_{k}(t)}{\sum\limits_{{k'=1}}^{k-1}{c^{i,j}_{k'}(t)p^{i,j}_{k'}(t) g^{i,j}_{k'}(t)}+ {I_{inter}(t) +\sigma^2(t)}},
 \end{align}
 where
\begin{align}
 I_{inter}(t)=\sum\limits_{i' \in \Phi_b\backslash b_i }{{\sum\limits_{  k' \in \Phi_u^{i',j}}}c^{i',j}_{k'}(t)p^{i',j}_{k'}(t) g^{i',j}_{k'}(t)}
\end{align}
and $p^{i,j}_k(t)$ is the transmit power of the user $u^{i,j}_k(t)$ and the set of transmit power is given by $\mathbf{P}_t$ $(p^{i,j}_{k}(t) \in \mathbf{P}_t, \forall i,j,k)$  \cite{8626185}. The power of thermal noise obeys $\sigma^2(t)=k_bT_rB$, where $T_r$ is temperature of resistors $k_b$ is Boltsmann's constant, $B$ is the considered bandwidth. In this paper we use $T_t=300$ K therefore, $\sigma^2(t)\approx 4.14 \times 10^{12} \: BW$. The $I_{inter}(t)$ represents the inter-cell interference, which is generated by the active users served by other BSs using the same sub-channel $s_j$.
In uplink NOMA, the decoding of user $u^{i,j}_k$ is based on the SIC process of its previous user $u^{i,j}_{k+1}$. If the data rate of successfully completing the SIC process is $R_{k+1}^{th}$, when the decoding rate of user $u^{i,j}_{k+1}$ obeys
\begin{align}
R^{i,j}_{k+1}(t) = \frac{B}{N_s}\log_2{\left(1+ \gamma^{i,j}_{k+1}(t) \right)} \ge R_{k+1}^{th},
\end{align}
the data rate of user $u^{i,j}_{k}$ is given by
\begin{align}
R^{i,j}_{k}(t) = \frac{B}{N_s}\log_2{\left(1+ \gamma^{i,j}_{k}(t) \right)}.
\end{align}
Otherwise, if $R^{i,j}_{k+1}(t) < R_{k+1}^{th}$, the decoding of all rest users $u^{i,j}_k, ... ,u^{i,j}_1$ fails, namely $R^{i,j}_{k}(t)=...=R^{i,j}_{1}(t)\equiv 0$.
\subsection{Problem Formulation}
For a long-term communication with period $T$, the number of active users is different across each time slot. Given the maximal load of each sub-channel $U_s$, we assume the number of active users are uniformly distributed in the range $[2,U_sN_bN_s]$ and $U_sN_bN_s \le N_u$. Under this condition, the average long-term sum rate can be maximized by optimizing clustering parameters $\mathbf{C} = \{\mathbf{C}_1,...,\mathbf{C}_T\}$ and transmit power $\mathbf{P} = \{\mathbf{P}_1,...,\mathbf{P}_T\}$. Therefore, the objective function is given by
\begin{subequations}\label{Rx}
\begin{align}\label {h}
\underset{\mathbf{C}, \mathbf{P} }{\max} \ \ \ & \frac{B}{N_s}\mathbb{E}\left[\sum_{t=1}^{T}\sum_{i=1}^{N_b}\sum_{j=1}^{N_s}\sum_{k=1}^{N_{u}^{i,j}}{\log}_2{\left(1+ \gamma^{i,j}_{k}(t) \right)}\right] ,\\
\mathrm{s.t:}\ \ \ \label {i}&g^{i,j}_1 \leq, ... ,\leq g^{i,j}_{{N_{u}^{i,j}}},\ \forall i, j, t,\\
&\label {j}{\sum_{k=1}^{N_u^{i,j}}{c^{i,j}_{k}(t) p^{i,j}_{k}(t) \le P_s ,\forall i, j, t}},\\
&\label {k}\gamma^{i,j}_{k}(t)  \geq 2^{R_k^{th}N_s/B}-1,\ \forall k, t,
\end{align}
\begin{align}
&\label {l} {2}\le\sum_{i=1}^{N_b}\sum_{j=1}^{N_s}\sum_{k=1}^{N_{u}^{i,j}}{c^{i,j}_{k}(t)\le N_u,} \ \forall t,\\
&\label {m} {\sum_{k=1}^{N_u^{i,j}}{c^{i,j}_{k}(t) \le U_s},}\ \forall i, j, t\\
&\label {o1}{\sum_{i=1}^{N_b}\sum_{j=1}^{N_s}c^{i,j}_{k}(t) \in \{1,0 \}} \ \forall k, t,
\end{align}
\end{subequations}
where \eqref{i} is the ordered channel gains based on the perfect CSI. \eqref{j} is to impose the power constraint of each sub-channel. \eqref{k} ensures all clustered IoT users can be successfully decoded for maximizing the connectivity. \eqref{l} and \eqref{m} limits the number of clustered users for the entire system and each sub-channel, respectively. \eqref{o1} indicates that each user belongs to only one cluster.
Problem \eqref{h} is an NP-hard problem, even only a fixed number of users per cluster is considered instead of dynamic range, especially, in case of \eqref{j} and \eqref{m}. The proof process is provided in Appendix A. The proof of \eqref{h} follows the idea in \cite{cui2018optimal} and \cite{8807386}.
\vspace{-0.1 cm}
\section{Intelligent Resource Allocation}
\subsection{Markov Decision Process Model for Uplink NOMA}
\begin{figure}
\centering
\subfigure[]{\label{sublable1}\includegraphics[scale=.3,keepaspectratio]{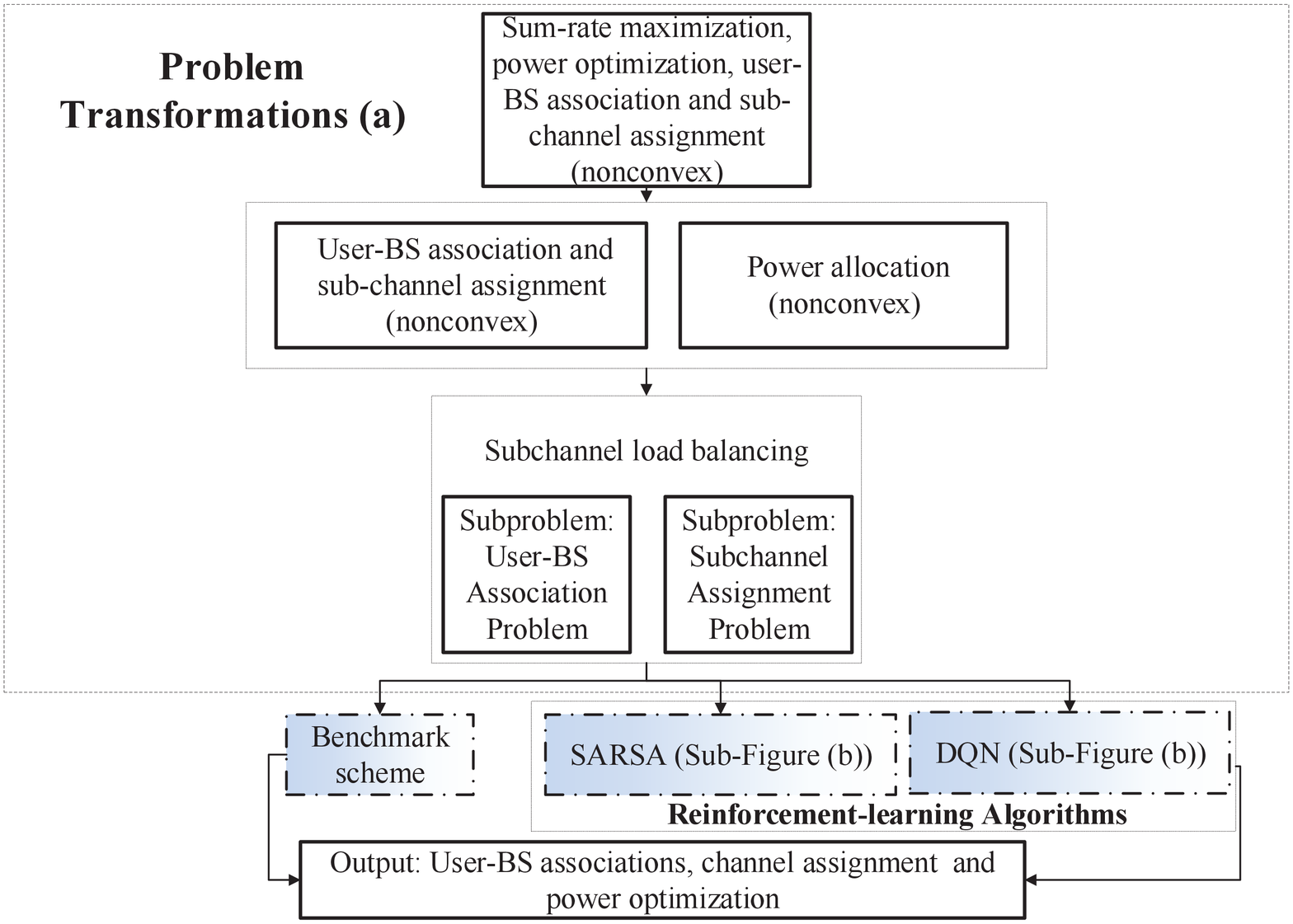}}
\subfigure[]{\label{sublable2}\includegraphics[scale=.3,keepaspectratio]{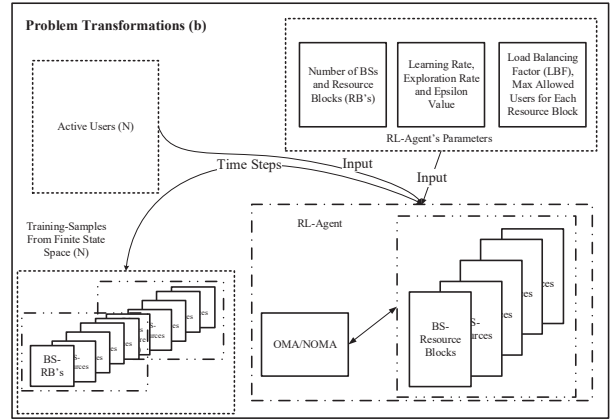}}
\caption{Overview of the proposed framework for the sum-rate maximization problem. Sub-figure (a) is an optimization problem breakdown to show where RL algorithms are applied and Sub-figure (b) shows problem transformations for the users and BSs as system states and the brain of reinforcement learning agents, respectively.}
\vspace{-0.9 cm} 
\end{figure}

In this section, we formulate user clustering and optimal resource allocation for uplink NOMA as a Markov decision process (MDP) problem. Problem transformations are shown in Fig. \ref{sublable1} and Fig. \ref{sublable2}. A general MDP problem contains single or multiple agents, environment, states, actions, rewards, and policies. The process starts with the interaction of an agent with a given environment. In each interaction, the agent processes an action followed by a policy $\pi$ with previous state $s$. After processing action according to these conditions and observed state agent/s receives a reward $r$ in the form of feedback to change its state $s^t$ to next state $s^{t+1}$. A reward can be positive (reward) or negative (penalty). It helps the agent/s to find an optimal set of actions to maximize the cumulative reward for all interactions. Q-table acts as the brain of an agent. The main function of Q-table is to store/memorize states $s$ and corresponding actions $a$ that the agent can take according to all the states as $Q^T_\pi(s,a)$ during trail $T$ for the basic RL algorithms.
SARSA and DRL are two promising RL methods to solve this MDP problem. SARSA learns the safest path, the policy $\pi^\prime$ is learned by estimation of state-value optimization function $Q^\prime(s, a)={Q}_\pi(s, a),\mathrm{}\forall s,a,$ but it requires more memory for complex state space. DRL uses a neural network to simplify the Q-table by reducing memory requirements to handle more complex types of problems. Furthermore, the design complexity for the SARSA algorithm is less because we only need to design Q-table. However, for the DRL algorithm, the design is more complex due to the deep neural network (DNN) and additional hyperparameters. Therefore, this work implements SARSA learning for light traffic. To further reduce the impact of state-space complexity DRL is used for heavy traffic scenarios. Additionally, in any case when the SARSA algorithm fails to provide an optimal policy for any type of network traffic during threshold trial $T_e$ then the final allocation is done using DRL.
 Finally, to summarize, this model follows model free on policy SARSA-learning algorithm instead of value iteration and off-policy methods for light traffic and DRL for complex networks. The major advantage of proposed algorithms is to avoid huge memory requirements (DRL) and learn the safest allocation policy (SARSA) for the different traffic conditions.
\vspace{-0.5 cm}
\subsection{SARSA-Learning Based Optimization For Light Traffic ((2-3,2-4)-Ue's)}
\begin{figure*}[t!]
\centering
\includegraphics[scale=.45,keepaspectratio]{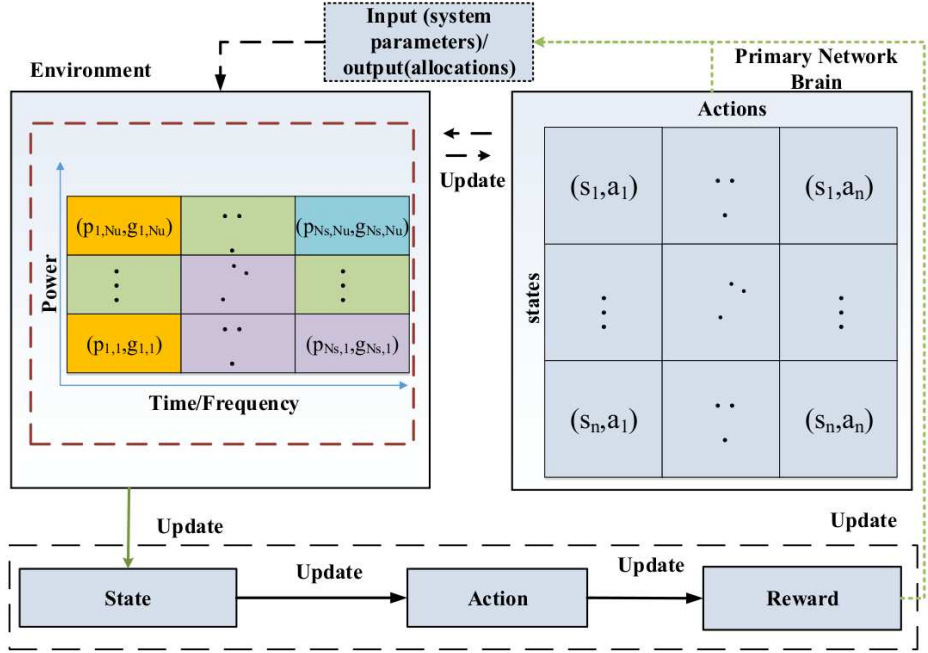}
\caption{An illustration of the communication environment for the proposed algorithm, where RL technique (SARSA) is invoked to optimize NOMA-IoT uplink 3D associations and resource allocation. The agents in this case are the BSs. The process of associations and resource allocation based on users activities is the state for our system.}
\label{sarsa}
\vspace{-0.5 cm} 
\end{figure*}

As the name suggests, for this type of traffic scenarios there is less number of users joining and leaving the network. In other words, the state space is not as huge as compared to heavy traffic. Therefore, we use the SARSA learning algorithm to find optimal long term policy. The traditional Q-learning is not suitable for the long term because it uses tuple of 3 $(S_t,A_t,R_t)$ for policy learning which doesn't know the knowledge of next step that is not suitable for our case. Secondly, the state space is not as huge as compared to heavy traffic that requires more complex control. To efficiently utilise system resources we use SARSA learning for light traffic and DRL for heavy traffic where the state space is huge with dynamic users. 
For SARSA learning, discount factor $\gamma$, sum reward, and the number of iterations are significant hyperparameters. The details for the flow of the information update is shown in Fig. \ref{sarsa}. The 5-tuple ($\mathbf {S}$, $\mathbf {A}$, $\mathbf {P}$, $\mathbf {R}$, $\mathbf {S^\prime}$, $\mathbf {A^\prime}$) SARSA-learning elements are mentioned below:
\begin{enumerate}
\item $\mathbf {S}$, is a state space consists of finite set having dimensions $N_b \times N_s$ containing $ N_u^{N_b \times N_s}$ total number of states. Each state represents one sub-set of 3D associations among users, BSs, and sub-channels.
\item $\mathbf {A}$, is an action space that consists of a finite set of actions to move the agent in a specific environment. Actions in this model are $[-1,0,+1]$. The '-1' is to reduce any one of the state elements from state matrix. Similarly, '+1' shows an increment in any of the state matrix elements. The last action '0' represents no change in the current state of the agent (BSs). It means that actions are swap operations between sub-channels and all BSs. For example, when an agent takes an action from (\ref{act}), the first action in $\mathbf{A}$ means agent performs swap operation of user between sub-channels at BS. In this model, agents have a total of $8$ swap operations between BSs and sub-channels.

\begin{align}\label{act}
  & \mathbf{A}= \left\lbrace \left( \begin{matrix}
   -1 & 0  \\
   1 & 0  \\
\end{matrix} \right),\left( \begin{matrix}
   1 & 0  \\
   -1 & 0  \\
\end{matrix} \right),\left( \begin{matrix}
   0 & -1  \\
   0 & 1  \\
\end{matrix} \right),\left( \begin{matrix}
   0 & 1  \\
   0 & -1  \\
\end{matrix} \right)\right., \nonumber \\ 
 & \left. \left( \begin{matrix}
   0 & 0  \\
   1 & -1  \\
\end{matrix} \right),\left( \begin{matrix}
   0 & 0  \\
   -1 & 1  \\
\end{matrix} \right),\left( \begin{matrix}
   1 & -1  \\
   0 & 0  \\
\end{matrix} \right),\left( \begin{matrix}
   -1 & 1  \\
   0 & 0  \\
\end{matrix} \right) \right\rbrace . 
\end{align}
\item $\mathbf {P}$, is an expected probability $P_{s\rightarrow s^\prime}^a=Pr{(}s^\prime|s,a)$ to change current state $s$ into next state $s^{\prime}$ by taking action $a$. The total number of actions for an agent are $(2 \times N_b \times N_s+1)$ with '8' swap operations. These operations include '+1','-1', and '0' actions, the agent selects suitable actions according to corresponding state to obtain an optimal state and action pair. 
\item $\mathbf {R}$, is a finite set of rewards, where the reward obtained after state $s$ transition to next state $s^{\prime}$ by taking action $a$. The reward function is denoted by $r_{s\rightarrow s^\prime}^a$, showing that in the result of all associations the agent will receive a reward according to the conditions mentioned in reward function.
\item Multi-constrained reward function, the short-term reward in the proposed model depends on two conditions:1) sum-rate and 2) the state of the system means the total number of users associated to BSs and sub-channels, which is defined as $\mathbf {S^\prime}$. The reward function can be expressed as follows:
\begin{align}\label{n}
r(s_t,s_{t+2},a_t)=
       \begin{cases}
       r=0,\text{$if \: R_{s_{t+1}} \geq R_{s_{t}}$}\\ \text{ and \ $ \sum_{t=1}^{T_e}(u_{k}^{s_{t}})= \sum_{t=1}^{T_e}(u_{k}^{s_{t+1}})$}\\
       r=-10,\text{otherwise.}
		\end{cases}
\end{align}
 \item $\mathbf {S^\prime}$, is a next state of an agent based on the previous state, action, and reward pairs of an agent.
 \item $\mathbf {A^\prime}$, is a next possible action can be taken by an agent from state $\mathbf {S^\prime}$.
 \end{enumerate}
\begin{definition}\label{def0.1}
The parameters of 3D state matrix $\mathbf {S}$ defined as $ Z=\lbrace{1,2,\cdots,N_u^{N_b \times N_s}\rbrace}$ total number of states with $N_b \times N_s$ dimensions $\sum_{1}^{Z}\mathbf {S}^{N_b}_{N_s}$. For all types of network traffic minimum for $Z$ is defined as $\sum_{1}^{N}Z_{(ij)} \geq 2$, the maximum for light traffic is $\sum_{1}^{N}Z_{(ij)} \leq \lbrace{3,4\rbrace}$ and for heavy network traffic the maximum load is $\sum_{1}^{N}Z_{(ij)} \leq 10$. 
\end{definition}
Furthermore, the optimal policy of the aforementioned parameters can be discovered by an agent using the following function:
\begin{align}\label {o}
    \pi^\prime(s)= {\mathop {\arg \max }\limits _{a}{Q}}^\prime(s,a),\forall s\in \mathbf {S^\prime},
\end{align}
where $\pi^\prime(s)$ represents the optimal policy. This function provides the optimal policy value for each state $s$ from the finite sate set after taking appropriate action $a$.
For a better understanding, the optimal policy can be defined:
\begin{align}\label {p}
    V_{\pi^\prime}(s_t)={\mathop { \max }\limits _{a}}{\left[r(s_t,a_t)+\gamma\sum_{s^\prime}{P_{s\rightarrow s^\prime}V_{\pi^\prime}(s^\prime)}\right]}.
\end{align}

For Q-table value updating that contains state and corresponding action values of an agent. Bellmen equation is utilised to perform optimization processes. According to the Bellman equation statement, there is only one optimal solution strategy for each environment setting. Bellman's equation is defined as:
\begin{align}\label {q}
    Q(s,a)\gets(1-a)Q(s,a)+\alpha\left[r^\prime+\gamma{Q}(s^\prime,a^\prime)\right],
\end{align}
where $\gamma \in (0,1)$ is a discount factor, which is a balancing factor between historical and future Q-table values. The larger $\gamma$ is the more weight for the future value and vice versa. $\alpha \in (0,1)$ indicates learning rate, it works like a step function (i.e., larger $\alpha$ contributes to fast learning but due to minimal experience, it may result in non-convergence. Similarly, if the value of $\alpha$ is too small then it will increase the time complexity of the system by leading it to a slow learning process).

\begin{definition}\label{def0}
For Q-learning we define $Q_{t=0}(s,a)=-100$ to learn greedy policy $\mathbf{P}_{\pi}(\mathbf{A}=a | \mathbf{S}=s)$  for all state and action pairs.   
\end{definition}
One main limitation of reinforcement learning algorithms is slow convergence due to $Q(s,a)$ requirement. Additionally, it is challenging with 3D state space and dynamic systems \cite{watkins1992q},\cite{melo2001convergence}. Due to dynamic behaviour of IoT users the 3D state $\mathbf{S}$ and action space, $\mathbf{A}$ influence learning process more as $\mathbf{S}$ and $\mathbf{A}$ are main parts of Q-table $Q(s,a)$. The convergence of the reward functions $r$ and reinforcement learning hyperparameters guide the algorithm towards optimal policy $V$. In other words, the choice of a reward function and the values for $\lbrace{\epsilon, \alpha \textit{ and, }\lambda\rbrace}$ are used by reinforcement learning agent/s to avoid the random walk. The random walk in search space causes infinite exploration of the search space resulting in no convergence. Therefore, we are able to propose the following conclusion.
\begin{remark}\label{remark1} The selection of suitable rewards $r_{s\rightarrow s^\prime}^a$  according to system dynamics $Q(s,a)$ is critical for effective convergence to find optimal $V_{\pi^{'}}$. Consequently, following \eqref{p} altering the reward function does not change the output of RL algorithms but the convergence towards policy $V_{\pi^{'}}$ is highly influenced.
\end{remark}
It is known that the proposed protocols are capable to handle multi-constrained optimization problems for different network traffic scenarios. We used $\epsilon-greedy$ SARSA-learning and DRL algorithms to explore and exploit search space to find dynamic outcomes, so the proposed protocols are capable to successfully obtain the optimal clustering solution. The Q-table in our model contains solutions for all subsets (user associations) in the search space. Therefore, in each episode $N_e$, only a specific subset of users will be active.
\begin{remark}\label{remark2}
In reinforcement learning to find the best associations $s_t$ from the set $S_t=\lbrace{s_1,s_2,\cdots,s_N}\rbrace$ possible states, an agent will converge towards the optimal states and actions pairing with the highest probability $\mathbf{P_{\pi^{'}}}$. In this way, by the increase in probabilities, the number of visits per state-action pair and rewards increase as well.
\end{remark}
Since an agent has limited successful visits, the achieved rewards will be as described in \textup{\textbf{Remark 1}} and \textup{\textbf{Remark 2}}. As a result, the agent successfully finds the optimal policy for the given system by processing the best actions. 
\subsubsection{SARSA-algorithm}
Based on the above discussions, we design \textbf{Algorithm~\ref{1}} for step by step significant optimization stages of the SARSA algorithm for light traffic networks. The details of the mentioned algorithm are as follows:
\begin{algorithm}[h!]
\tiny
\caption{SARSA-Learning Based NOMA-IoT Uplink Resource Optimization }
\label {1}
\begin{algorithmic}[1]
\State Inputs for SARSA:
\begin{enumerate}
\item Episodes $N_e$
\item Explorations per trials $T_e$
\item Learning rate $\alpha$
\end{enumerate}
\State Initialization for SARSA:
\begin{enumerate}
\item Network parameters ($N_b,b_i,N_s,s_j,N_u,u_k,P_b$)
\item Q-Table $Q(s,a)$
\end{enumerate}	
\State Define number of clusters-k
    \State Define range of users per cluster
    \State load $\textbf{s}= N_u^{N_b \times N_s}$ and $\textbf{a}=[-1,0,+1]$
    \State Random user association to any $BSs$ and Cluster
    \For {iteration = $1$:$N_e$}
    \State  $s_t=rand()$
        \For {iteration = $1$:$T_e$}
        	\State $s_t,a_t$
        	\State compute $   r(s_t,s_{t+2},a_t)=
       \begin{cases}
       r=0,\text{$if \: R_{s_{t+1}} \geq R_{s_{t}}$}\\ \text{ and \ $ \sum_{t=1}^{T_e}(u_{k}^{s_{t}})= \sum_{t=1}^{T_e}(u_{k}^{s_{t+1}})$}\\
       r=-10,\text{otherwise.}
		\end{cases}$              	
        	\State update $ R_{sum}$       				          		       				        						
        	\State update  $Q(s,a)\gets(1-a)Q(s,a)+\alpha\left[r^\prime+\gamma{Q}(s^\prime,a^\prime)\right].$
        	\State Update $\pi$ towards greediness
        	\State  $s \leftarrow s^\prime, a \leftarrow a^\prime$       	
		\EndFor		
   	        \State return optimised \textbf{(c,p)} (\ref{Rx}) under constraints (\ref{h}),(\ref{i}),(\ref{j}),(\ref{k}) and (\ref{l})   	         	
   \EndFor
	\State Return Q-Table $Q(s,a)$	
\end{algorithmic}
\end{algorithm}
\begin{itemize}
\item Line $\#(1-6)$: presents the initialization of the SARSA algorithm, in which the system is initialized by initial sets of users, BSs, and sub-channels as an initial state $S_t$. After this, we define the maximum number of clusters and the maximum number of users for each cluster. In line$\#2$ the brain of an agent is initialized with $-100$ having dimensions $[s \times a]$ as Q-table. The purpose of initialization with $-100$ is to show that the brain of an agent needs training. Therefore, after training, the Q-table will contain values approaching zero for the best case and vice versa. Secondly, it also shows that the proposed algorithm is targeted to solve the maximization problem, maximum Q-value means better solutions. Line $\#(3-6)$ shows SARSA-learning parameter definition and initial random association among IoT users, BSs, and sub-channels.
\item Line $\#(7-17)$: shows key training steps based on Q-table updates via bellman equations. From line$\#1$, an agent performs actions according to a given state of the environment, that is 3D associations and cluster allocation. In line$\#8$ agent picks new associations for different active users in one episode, then for all trials agent is trying to get optimal associations with optimal sum rate, if the associations are successful then the agent gets a reward (0) and if it fails then negative (-10) is given as a punishment. In line$\# 13$, based on the 3D designed 5-tuple ($\mathbf {S}$, $\mathbf {A}$, $\mathbf {P}$, $\mathbf {R}$, $\mathbf {S^\prime}$, $\mathbf {A^\prime}$) values (\ref{q}) is updated on-line. To perform online updates using $\mathbf {S}$, $\mathbf {A}$, $\mathbf {P}$, $\mathbf {R}$, $\mathbf {S^\prime}$, $\mathbf {A^\prime}$ instead of $\mathbf {S}$, $\mathbf {A}$, $\mathbf {P}$, $\mathbf {R}$ as (traditional Q-learning) the online learning mechanism becomes more fast converging. In other words, the agent finds optimal long-term online allocation policy more efficiently. Similarly, these updates are calculated for maximum episode $N_e=500$ and all the trials $T_e=500$ to maximize the overall long-term average reward of the system.
\end{itemize} \begin{definition}\label{def0.11}
In 3D state matrix $\mathbf {S}$ from the set $S_t=\lbrace{s_1,s_2,\cdots,s_N}\rbrace$ possible states, is defined as CSI of the proposed network that is known to both of the reinforcement learning agents. Therefore, the reinforcement learning agent contains perfect knowledge of the CSI for the whole network.  
\end{definition}
\subsection{Deep Reinforcement Learning For Heavy Traffic ((2-10)Ue's)}
In general, both on-line and off-line Q-learning methods require high memory space to build a state of the systems. However, practical systems are high in dimension and complex. Due to this reason Q-learning is not suitable for a large action space, this is a major drawback of conventional Q-learning methods. To overcome this, DRL method adopts a DNN ${Q}(s, a;\theta)$, to generate its Q-table with the help of $\theta$ by approximating the Q-values ${Q}(s,a)$  \cite{silver2017mastering}. Therefore, DRL agents only need to memorize the $\theta$ weights instead of reserving huge memory space for all possible states and action pairs. This is the main advantage to use DNN. More specifically, in conventional Q-leaning algorithms, the optimization of ${Q}(s,a)$ is equal to the optimization of ${Q}(s,a;\theta)$ in DRL with low memory requirements. Similarly, $\theta$ updates are based on history states, actions, and reward values. More specifically, these values are based on DRL agent interactions with the environment to learns the relationship among the different actions and states by continuously observing a given environment. 
\begin{enumerate}
  \item $\mathbf {S}$, is a unique state space used as an input of DNN. Each state is a combination of multiple sub-sets of 3D associations among users, BSs, and sub-channels. It also consists of current rewards of the system as an instantaneous and average reward from previous iterations. 
  \item $\mathbf {R}$, is a reward of the system that is denoted by $\mathbf {R}= \lbrace{r_{i},r_{l}}\rbrace$, where $r_{i}$ is an instantaneous rewards similar to SARSA algorithm and $r_{l}= \sum_{t=1}^{500}{r_{i}^{t}/t}$ denotes long-term average rewards for the time slot $t$.  
  \item $\mathbf {A}$, is a multi-dimensional matrix representing actions as $\mathbf {A}= \lbrace{\mathbf{a_{1}},\mathbf{a_{2}},\cdots,\mathbf{a_{8}}}\rbrace$. For the DRL algorithm, the action mechanism is based on two main parts as; allocation strategies described as switching strategy $\mathbf {a_{s}}$ and association strategy $\mathbf {a_{i}}$ for the optimization process, where $\mathbf {a_{s}}$ is a switching mechanism similar to SARSA and used for the DRL channel switching process. The second strategy $\mathbf {a_{i}}$ is a result of selected switching strategy $\mathbf {a_{s}}$, $\mathbf {a_{i}}$ denotes an index of the 3D associations among users, BSs, and sub-channels. Finally, the DRL agent uses loss function mentioned in (\ref{dqn1}) to calculate $\theta$ based on the previous experience. 
\begin{align}\label{dqn1}
loss(\theta) = 1/N_e \sum_{t=1}^{N_e}{{{[y^{DRL}_t - Q(s_t,a_t;\theta)] }^{2}}}, 
\end{align}
where
\begin{align}\label{dqn2}
y^{DRL}_t = r + \gamma \max \limits _{a' \in \boldsymbol {A}} ~Q(s',a';\theta^{'})
\end{align}
and $y^{DRL}_t$ is the target Q-values from target DNN. For the improved training, in general the update frequency of the target network $\theta^{'}$ is performed in slow manner. Due to this reason the target network remains fixed for the target network update threshold $T_e^{'}$.
\end{enumerate}  
The DRL agent uses gradient decent method as in (\ref{dqn1}) to reduce the prediction error by minimizing the loss function. The updating of $\theta$ is provided in (16), which is based on the outcome of new experience. The updating function for $\theta$ is defined in (\ref{dqn7}), namely DRL Bellman equation.
\begin{align}
&\label{dqn3}{\theta \gets \theta -[ y^{DRL}} - Q(s,a;\theta)]\nabla Q(s,a;\theta).\\
&\label{dqn4}q_{\pi }(s, a) = r(s, a) + \gamma \sum \limits _{s' \in \boldsymbol {S}} \sum \limits _{a' \in \boldsymbol {A}} {{p_{ss'}}(a)} q_{\pi }(s', a'), \\
&\label{dqn5}q_{\pi ^{*}}(s,a) = r(s,a) + \gamma \sum \limits _{s' \in \boldsymbol {S}} {{p_{ss'}}(a)\max \limits _{a' \in \boldsymbol {A}} } ~q_{\pi ^{*}}(s',a'),
\end{align}
where the function $q_{\pi ^{*}}(s,a)$ shows Q-values and the long-term reward calculations for DRL based on the discount factor $\gamma$ and below mentioned optimal DRL policy $\pi ^{*}$.
\begin{align}\label{dqn6}
{\pi ^{*}}(s) = \mathop {\arg \max }\limits _{a \in \boldsymbol {A}} \left [{ {q_{\pi ^{*}}(s,a)} }\right], ~\forall ~s\in \boldsymbol {S},
\end{align}
where $\pi^{*}(s)$ represents the optimal policy for the DRL algorithm. This function provides the optimal policy value for each state $s$ from finite sate set after taking appropriate action $a$.
\begin{align}\label{dqn7}
Q(s,a) \!\gets \!(1\!-\!\alpha)Q(s,a) \!+\!\alpha \!\left [{\! {r(s,a) \!+\! \gamma \max \limits _{a' \in \boldsymbol {A}} Q(s',a')} \!}\right]\!,\quad
\end{align}
where $Q(s,a)$ is showing Q-value update according to DRL Bellman equation.
\begin{align}
&\label{dqn8}s(t)=\{a_{1}(r^{1}_{i},r^{1}_{l}),a_{2}(r^{2}_{i},r^{2}_{l})\ldots, a_{t}(r^{t}_{i},r^{t}_{l})\}. \\
&\label{weights} \rho(x):\sum_{j=0}^{d_j} W_j \times I_j(s(t)) + \psi_j,
\end{align}
where in \eqref{dqn8}, $s(t)$ represents state of the DRL agent and equation (\ref{weights}) shows the activation mechanism for each neuron layer $I$ based on weights $W_j$ for $j-th$ depth of neurons with bias term $\psi_j$. In this model, the input of the DRL algorithm is the instantaneous network observation as $s_t$. This state is sent to the different neural network neurons with specific network $W_j$ to obtain the final output as a set of different Q-values for all actions. For the DRL framework, the size of output actions is similar to the SARSA.
We use the replay memory as an experience for the DRL agent to store the tuple $( s_t,a_t,r_t,s^{'})$ for all the time steps $T_e$ in an experience dataset $E$ with size $\varepsilon$. When the size $\varepsilon$ is full, the first experience as the oldest tuple will be removed to free some space for the new experience update. The reason for this updated experience is to reflect the sequential exploration of the DRL framework. However, the distribution of the samples is independent and identical. Therefore, to get more general output, the $W_j$ update process is performed based on randomly sampled tuple $( s_t,a_t,r_t,s^{'})$ instead of the current tuple. This is because output is highly influenced by the correlated set of tuples $( s_t,a_t,r_t,s^{'})$ and variance of the updates.
\begin{definition}\label{def1.0}
DRL design in this work is defined with two main elements, the first element is target Q-network based on $\theta^{'}$. The second main element of DRL is state transition mechanism $( s_t,a_t,r_t,s^{'})_{t \in [n]}$. This mechanism is used to construct mini-batch for experience reply from dataset $E$ to train DNN.   
\end{definition}   
\begin{figure*}[htp!]
\centering
\includegraphics[scale=.5,keepaspectratio]{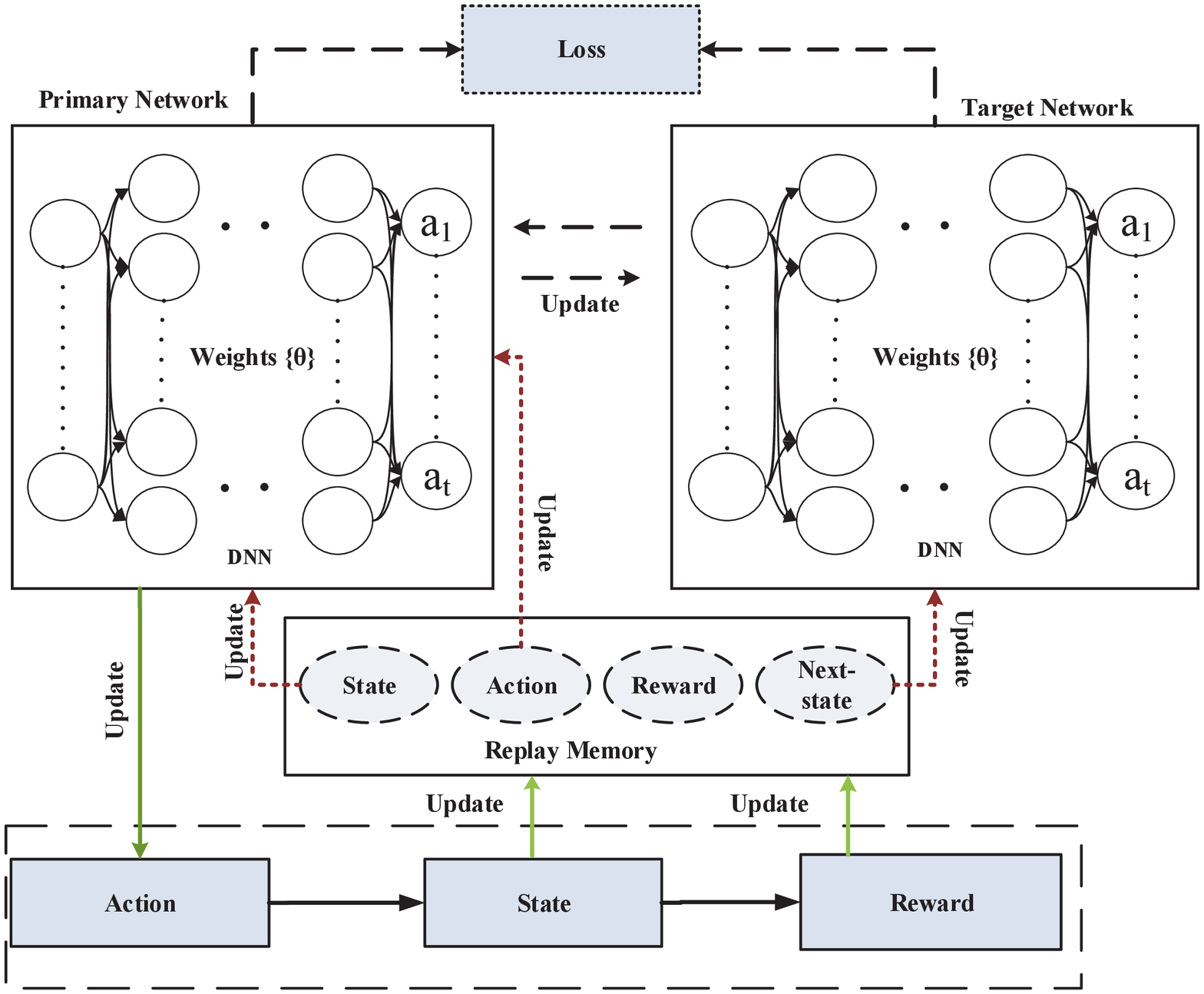}
\caption{DRL structure: it shows the flow of information between target and training networks to minimize the loss function using states, actions, rewards, and replay memory.}
\label{drl}
\vspace{-0.5 cm} 
\end{figure*}
\begin{remark}\label{remark3}
The convergence rate/speed of the proposed algorithm varies according to the initial 3D association (states) that is randomly selected. In this model, the state space means allocation strategies that include subsets of all possible associations of active users $u_k = 2 \leq N_u$ for each sub-channel at the episode $N_e$.
\end{remark}
Based on the above discussions, we design \textbf{Algorithm~\ref{11}} for step by step significant optimization stages of the DRL algorithm for heavy traffic. The details of the mentioned algorithm are as follows:
\subsubsection{DRL}
\begin{itemize}
\item Line $\#(1-2)$: In this stage, the parameter initialization is performed, which is a similar initialization step like SARSA. However, instead of state action pairing, the weight matrix is initialized for DRL to find optimal policy $\pi$.
\item Line $\#(3)$ Pre-training: In this stage, initial actions are selected using uniform random distribution as an initial state space in a continuous environment. In this way, initial weights are also calculated to start the optimization process.
\item Line $\#(4-17)$: Whole process for DRL is similar to SARSA from line $\#(5-11)$ with DRL bellmen equations (\ref{dqn1}) to (\ref{dqn8}). 
\item Loss Calculation: The equation \eqref{dqn1} is to calculate the loss $\theta$ that is the mean squared error (MSE) indicating the difference of the target and predicted networks. To optimise these values between the target network and prediction network we use Adam optimiser. The Adam optimizer is used for the loss minimization to further improve the optimal predicted Q-values for each episode. Therefore, the DRL framework converges faster even in huge state space. In \eqref{dqn2}, we calculate the target Q-values based on the tuple $(S_t,A_t,R_t)$ from mini-batch and the mini-batch is updated after 100 iterations.  
\item DRL Updates: The updating function for the prediction of DRL $\theta$ and long-term reward calculation is shown in \eqref{dqn3} to \eqref{dqn5}, where DRL agents obtain rewards and prediction loss after every transition from $s_t$ to next state $s_{t+1}$ to find the greedy policy. Additionally, $\gamma$ discount factor has a significant impact in search of the greedy-policy because based on discount factor as we mentioned in the previous paragraphs, the agent selects immediate or previous Q-values. The policy $\pi$ is calculated using \eqref{dqn6} to maximize the Q-values by the greedy search. The calculation for DRL Bellman equation is performed using states in \eqref{dqn7}. 
\item Sparse Activations: The $\rho(x)$ is an activation function for DNNs sparse activations using ReLU $(\rho)$. The sparse activations help agents to efficiently converge by avoiding useless neuron activations. The outcome of sparsity is shown in the results section, comparing sparse ReLU, Sigmoid, and TanH. In \eqref{weights}, the activations are performed for the $\delta_j$ density of neurons with $j-th$ index, for each neuron we use state of the system as an input that is multiplied with weight $W_j$ of $j-th$ density and adding bias value as $\lambda$ before activation. In the next steps, current states, actions, and rewards are added to mini-batch for experience replay (for self-training). In $\#(10-14)$, the agent receives next states from mini-batch that is learned in previous sections based on pre-training. Before that, the learning process of the agent is based on pre-training but when mini-batch is full, the agent will learn the optimal policy by experiencing a replay mechanism with the help of mini-batch processing.
\item  Neural Networks: this paper uses the DRL that is built with two DNNs as shown in Fig. \ref{drl}: 1) a training network ${Q}(s,a;\theta)$ that learns the policy and 2) a target network ${Q}(s',a';\theta')$ to compute target Q-values for every update, where $\theta$ and $\theta'$ shows the weights of these two networks. For the training of the DNN network, $\theta$ weights are predicted based on the current state and action. On the other hand, $\theta'$ weights are based on the previous episodes and these weights are fixed during the calculation of $\theta$ for training purpose. Additionally, We utilize MSE loss function (\ref{dqn1}) to evaluate the accuracy of the training for the target network. Therefore, the proposed loss function is based on $\theta$ and $\theta'$ to check the deviation of the predicted DNN weights.
\item  Output: Finally, the output of this algorithm is the optimal policy for all clusters where the overall long-term sum rate is maximum.
\end{itemize}
\begin{definition}\label{def1}
We use ReLU activation function $\rho(x)= max(x,0)$ ($x$ is the input neuron) for DRL performance evaluations. A ReLU network of density $\delta_j$ and $\lambda_i$ hidden layers with each layer width $\lbrace \delta_j \rbrace^{\lambda+1}_{i=0} \subseteq \mathbb{N} $ can be represented as $f: \mathbb{R}^{\delta_0} \rightarrow \mathbb{R}^{\delta{\Lambda+1}}$ for any positive number $L$. $f(x)={{w}_{\Lambda+1}}\rho ({{w}_{\Lambda}}\rho ({{w}_{\Lambda-1}}...\rho ({{w}_{2}}\rho ({{w}_{1}}\rho +{{\psi}_{1}})+{{\psi}_{2}})$ $ ...{{\psi}_{\Lambda-1}})+{{\psi}_{\Lambda}}.$
\end{definition}
\noindent In this definition, $f(x)$ is a function to show the construction of neural network with weights for each layer ${w}_{\lambda} \in \mathbb{R}_{\delta_\lambda}^{\delta_{\lambda-1}}$ and $\rho$ is the activation for each neuron. The mesh structure of the neural network remains fixed in this model to learn two main neural network parameters $({w}_{\lambda},{\psi}_{\lambda})_{[\lambda \in \Lambda+1]}$ in addition with the activation function $\rho$ and the input of the neural network. In the neural network $\Psi$ bias terms are added with the input of the DNN as $\Psi_{\Lambda+1}$ as a shift value.   
To optimise our dynamic objective function, the greedy search agent is used. With the help of a greedy search, the DRL agent receives higher rewards.
\begin{remark}\label{remark4}
 To avoid useless visits, greedy policy $\pi^{'}$ provides a balanced exploitation, because $\epsilon-greedy$ exploits in the most cases and some times it processes a random action to explore the environment in search of different solutions $Q_{N_e}(s,a)= \mathbb{E}[\sum_{t=1}^{N_e}{\gamma_{t}r_t}]$. 
\end{remark}
For DRL, unbalanced random actions cause huge error propagation so that this $\epsilon-greedy$ is suitable to be applied for achieving efficient learning in dynamic state space. Note that the boundary for the policy selection is $0\leq \epsilon \leq 1$. For $\epsilon$ close to $0$ the policy becomes a greedy policy, and for $\epsilon$ close to $1$ the agent explores more.  
\begin{algorithm}[htbp!]
\tiny
\caption{Deep Q-Learning Based NOMA-IoT Uplink Resource Optimization  }
\label{11}
\begin{algorithmic}[1]
\State Inputs for DRL:
\begin{enumerate}
    \item Episodes $N_e$
    \item Explorations per trials $T_e$
    \item Learning rate $\alpha$
\end{enumerate}
\State Initialization for DRL:
\begin{enumerate}
    \item Network parameters ($N_b,b_i,N_s,s_j,N_u,u_k,P_b$)
    \item memory, hidden size, State size, action size and mini-batch
\end{enumerate}	
\State train DRL to find a good policy $\theta$ 	
\For {iteration = $1$:$N_e$}
  \For {iteration = $1$:$T_e$}
       \State $s_t,a_t$
       \State compute $   r(s_t,s_{t+2},a_t)=
       \begin{cases}
       r=0,\text{$if \: R_{s_{t+1}} \geq R_{s_{t}}$}\\ \text{ and \ $ \sum_{t=1}^{T_e}(u_{k}^{s_{t}})= \sum_{t=1}^{T_e}(u_{k}^{s_{t+1}})$}\\
       r=-10,\text{otherwise.}
		\end{cases}$
       \State update $\theta$  using  $q_{\pi ^{*}}(s,a) = r(s,a) + \gamma \sum \limits _{s' \in \boldsymbol {S}} {{p_{ss'}}(a)\max \limits _{a' \in \boldsymbol {A}} } ~q_{\pi ^{*}}(s',a').$
       \State $loss(\theta) = {{{[y^{DRL}_t - Q(s_t,a_t;\theta)] }^{2}}},$ update using $y^{DRL}_t = r + \gamma \max \limits _{a' \in \boldsymbol {A}} ~Q(s',a';\theta').$
       \State  $s \leftarrow s^\prime, a \leftarrow a^\prime$
       \State update mini-batch (Experience)
  \If {$T_e > State-size$ }
 	\State get $s \leftarrow s^\prime, a \leftarrow a^\prime$ from mini-batch
 \EndIf
  \EndFor   	
\EndFor
\State Return $Q(W_a)$
\end{algorithmic}
\end{algorithm}
\begin{definition}\label{def1.1}
(Sparsity for ReLU DNN): The sparsity of the ReLU network is a weight based sparsity denoted by $\kappa$, sparse ReLU networks are bounded by $\Psi$ for $\Lambda_i$ layers, $\Psi > 0$ . For any hidden layer $\Lambda_i,\kappa \in \mathbb{N}, \lbrace{\delta_j}\rbrace_{i=0}^{\Lambda+1}\subseteq \mathbb{N}$.
\begin{align*} 
\mathsf{F}(\Lambda,\{{{\delta}_{j}}\}_{i=o}^{\Lambda+1},\kappa,\Psi)= \left( f:\underset{\lambda \in [\Lambda+1]}{\mathop{\max }}\,{{\left\| {{W}_{\lambda}}^{'} \right\|}_{\infty }}\le 1, \right.
\end{align*} 
\begin{align*}
\left. \sum\limits_{\lambda=1}^{\Lambda+1}{{{\left\| {{W}_{\lambda}}^{'} \right\|}_{0}}\le \kappa,\underset{j \in [\delta_{\Lambda+1}]}{\mathop{\max }}\,{{\left\| {{f}_{j}} \right\|}_{\infty }}\le \Psi} \right),
\end{align*}   
\end{definition}
\noindent where ${W}_{\lambda}^{'}$ is used to represent ${W}_{\lambda},{\psi}_{\lambda}$. The function $f$ is from \textbf{Definition \ref{def1}} and $f_j$ is the $j-th$ element of $f$.
\vspace{-0.5 cm} 
\subsection{Complexity}
\vspace{-0.2 cm} 
The complexity of the proposed model is based on the number of BSs $N_b$, total number of sub-channels $N_s$ and the number of users communicating $N_u$. In proposed scheme, simulated experiments are based on different examples. This paper considers $N_b=2$, $N_s=2$ and $N_u=3,4,$ for light traffic and $10$ for heavy traffic. These examples are association decisions for the user $N_u$ and the sub-channel $N_s$ at BS $N_b$ that receives signals for $N_u^{N_b}$ channels from users. The computation complexity for SARSA-learning is $\mathcal{O}(N_bN_u^{N_s})$ operations with $N_u^{N_b \times N_s} \times (2 \times N_b \times N_s+1)$ memory requirement for Q-table to simulate brain of the learning agent/s. The complexity of DRL is $\mathcal{O}(N_eT_e)$ with smaller Q-table $\mathcal{O}(Q(W_j))$ and DRL uses 1D experience replay containing states vector (\ref{dqn8}) instead of huge memory requirements like traditional Q-learning. The benchmark scheme considered in this work is a memory-less method, which shows the maximum achievable rate by exhaustively searching all possible combinations of 3D associations. Consequently, it requires more number of operations. Due to this reason, the computation complexity increases in exponential manner as $\mathcal{O}(N_u^{N_b \times N_s})$.
\begin{table*}[htb!]
\scriptsize
\centering
\caption{Network parameters}
\label{tab2}
\begin{tabular}{|l|l|l|l|}
\hline
\hline
   \text{Load balancing factor $k$ values per resource block} & 2-3,2-4,2-10& \text{Total number of trials} &  500  \\
   \text{Total number of time steps} &  500&  \text{Bandwidth} &  $[15-35]kHz$  \\
   \text{Gain} &  $[1,1.5,2]*10^{-5}$ \cite{zhang1999finite}&    \text{$\gamma_{\delta}$}  &$0.6$  \\
   \text{$\alpha$} & $0.75,0.1$&   \text{$\epsilon$} & $0.1$   \\
   \text{$\lambda$} & $0.5$ &     \text{Optimisers} & SARSA-DRL (Adam)   \\
   \text{Deep neural network activations} &  Sigmoid, TanH, ReLU & Power levels&[5,10,15,20,25,30]dBm  \\
\hline
\hline
\end{tabular}
\end{table*}

\section{Numerical Results}
\vspace{-0.2 cm} 
  In this section, simulation results are provided for the performance evaluation of the proposed multi-constrained algorithms. The proposed multi-constrained algorithms are tested under different network settings to solve: 3D associations among user, BSs, and sub-channels as well as sum-rate optimization with different network traffic.  For simulations, we have considered two different traffic density threshold values to analyse the impact of network load with various power levels on the sum rate and 3D associations. Additionally, the network load in our case represents the load of each resource block instead of the total number of users in the network. Therefore, max network load=10 with two RB's for each BSs means $10*4=40$ users in the network. To show the significance of available channel bandwidth, we start with a minimum channel bandwidth of 60(kHz) and then increase it to 120(kHz) under different network traffic conditions. The hardware and software system used for experimentation is Intel core i7-7700 CPU with 3.60 GHz frequency having 16 GB of RAM (Random Access Memory) and 64-bit operating system (windows-10). All the experiments are simulated using Matlab version-R2019a and Python 3.6.
 From Table.\ref{tab2} for both the algorithms we used $500$ episodes with $500$ iterations for each episodes. Similarly, $\lambda_{[SARSA/DRL]} =0.5, \gamma_{[SARSA/DRL]} =0.6, \alpha_{[SARSA/DRL]} =\lbrace{0.75,0.1}\rbrace,$ and $\epsilon-greedy$ exploration are values for the significant hyper-parameters of proposed algorithms. We used Load balancing factor $k$ values per resource block to show the maximum and minimum user connectivity for each resource block. The values of channel gain for each user is defined as $[1,1.5,2]*10^{-5}$ \cite{zhang1999finite}. For the DRL, additional parameters are trained, such as loss MSE, activation functions, batch-size$=500$, optimisers, experience memory $E=500$, pre-training length $=500$, the number and size of hidden units. We use ReLU, Sigmoid, and TanH as activation functions with two hidden layers having density $\delta=500$ units. Adam optimizer is utilized for the optimal convergence of DRL.    
\begin{figure}[H]
\centering
    \subfigure[]{\label{con}\includegraphics[scale=.5,keepaspectratio]{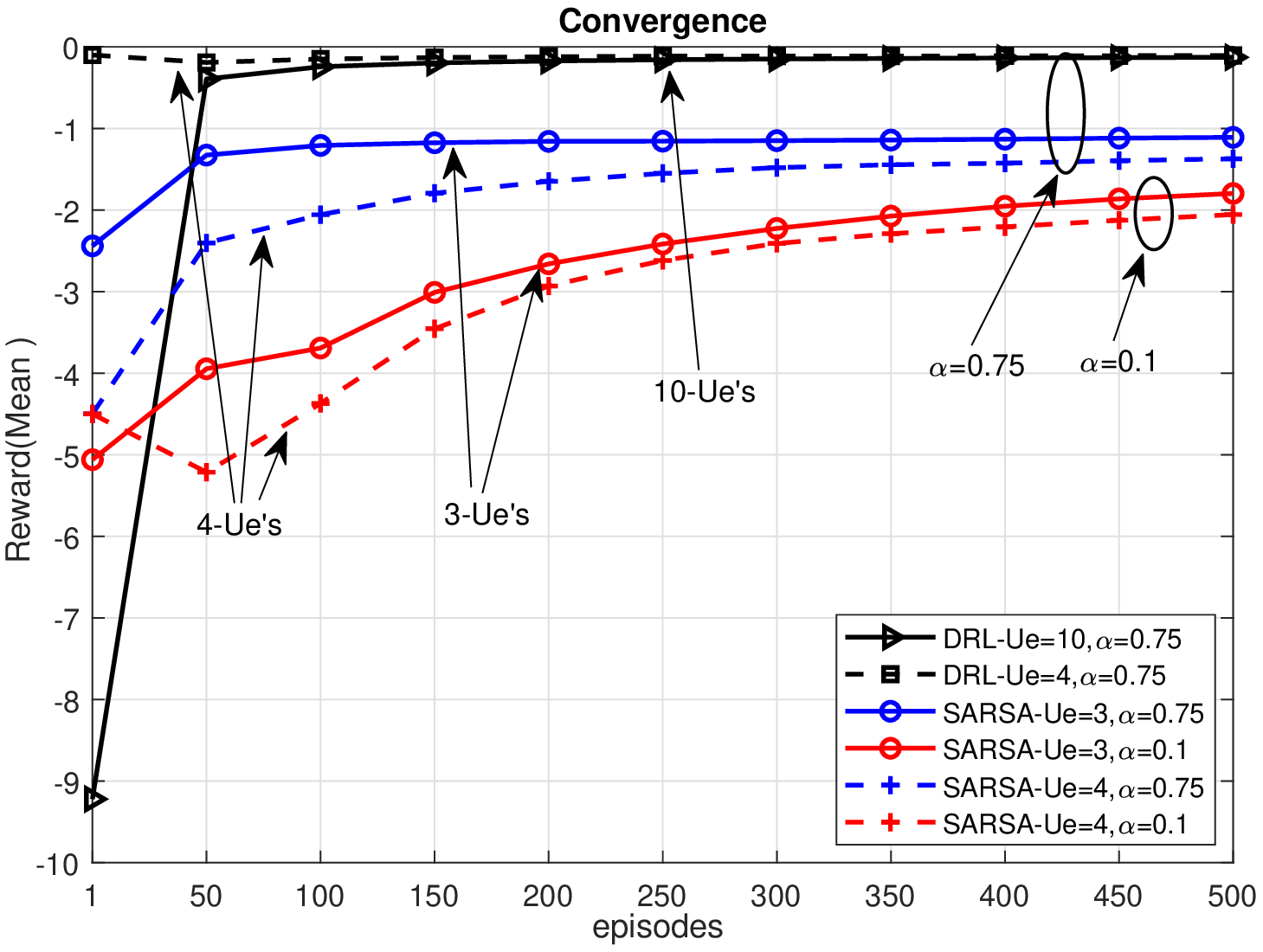}}
    \subfigure[]{\label{longb}\includegraphics[scale=.5,keepaspectratio]{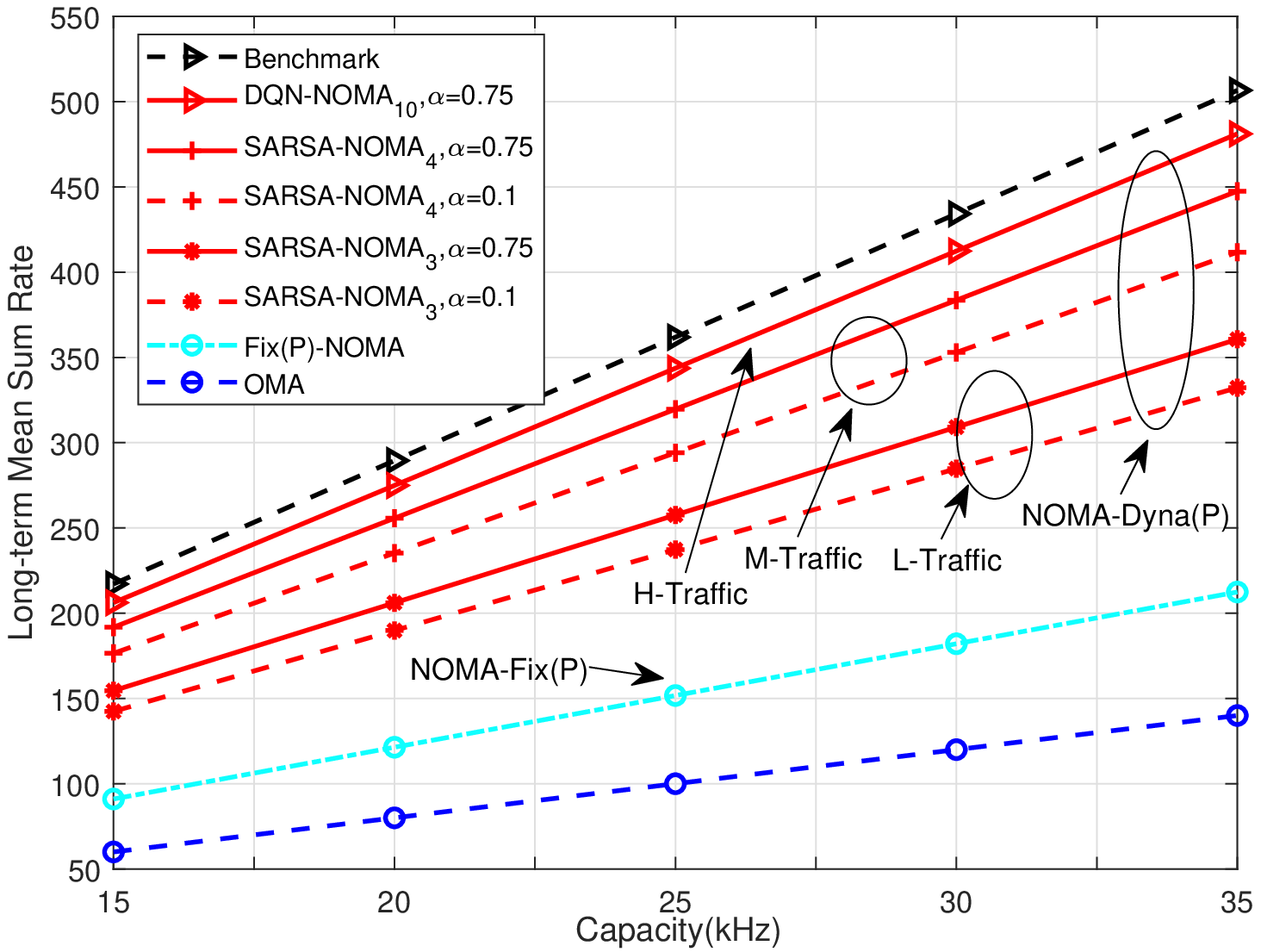}}
\caption{ Sub-figure (a) is the convergence for the proposed algorithms: DRL for medium and heavy traffic (for 4 users and max scheduling up to 10 users), SARSA for medium and low traffic range ( support 2, 3, and upto 4 users scheduling) and the comparison for two different learning rates ($\alpha=0.75, \alpha=0.1$). The sub-figure (b) shows a long-term comparison between the channel bandwidth, average sum rate, and power levels for the proposed SARSA, DRL and benchmark scheme.}
\vspace{-0.5 cm} 
\end{figure}
\vspace{-0.2 cm} 
\subsection{Convergence vs Sum Rate vs Traffic Density}
Fig. \ref{con} shows the inter-correlations among the four measures of convergence. It is apparent from this figure that if traffic density increases then convergence is slower and vice versa. DRL has better convergence for heavy traffic with the maximum allocation capacity/load, which makes DRL more suitable for scenarios with high traffic densities. Furthermore, to show the performance of DRL with medium traffic (M), we can see that compared to the SARSA algorithm DRL is handling medium traffic better by fast converging. Secondly, another interesting insight is that the performance of SARSA $\alpha=0$ is better than $\alpha=0.1$ with $\epsilon-greedy$. The convergence of Adam depends on DRL $\theta$ weights as $R_{DRL}=\sum\limits_{t=1}^{T}{\left( {{f}_{t}}({{\theta }_{t}})-{{f}_{t}}({{\theta }^{'}}) \right)}$. where ${\theta }^{'}= \mathop {\arg \min}_{\theta \in \kappa} \sum_{t=1}^{T}f_t(\theta)$ and $\kappa$ is feasible set for all $t-1$ steps.
\begin{definition} The bounded gradients of the function $f_t^{DRL}$ is ${{\left\| \delta {{f}_{t}}(\theta ) \right\|}_{2}}\le {{G}^{DRL}}{{\left\| \delta {{f}_{t}}(\theta ) \right\|}_{\infty }}\le G_{\infty }^{DRL},\forall \theta \in R_{d}^{DRL}$. Secondly, the distance generated by the Adam optimiser is bounded as:  ${{\left\| {{\theta }_{p}}-{{\theta }_{q}} \right\|}_{2}}\le D,{{\left\| {{\theta }_{p}}-{{\theta }_{q}} \right\|}_{\infty }}\le {{D}_{\infty }}$ for any $ p,q \in {\lbrace 1,\cdots,T}\rbrace$ with the bias terms $\beta_1,\beta_2 \in [0,1]$ satisfying the $\frac{\beta _{1}^{2}}{\sqrt{{{\beta }_{2}}}}<1$ condition. Let the learning rate of the Adam optimiser be $\alpha_t=\alpha/\sqrt{t}$ and bias term $\beta_1^t= \beta_1\lambda^{t-1},\lambda \in [0,1]$ for each step $t$, for all $T \geq 1$ Adam obtains the following condition \cite{kingma2014adam}:
\end{definition}
\vspace{-1cm}
\begin{align}
  &{{R}_{DRL}}(T)\le {{D}^{2}}/2\alpha (1-{{\beta }_{1}})\ \sum\limits_{i=0}^{d}{\sqrt{Tv_{T,i}^{'}}} \nonumber \\
  &+\frac{\alpha (1+{{\beta }_{1}})G_{\infty }^{DRL}}{(1-{{\beta }_{1}})\ \sqrt{1-{{\beta }_{2}}}{{(1-\gamma )}^{2}}}\sum\limits_{i=0}^{d}{{{\left\| {{g}_{1:T,i}} \right\|}_{2}}} \nonumber \\
 & +\sum\limits_{i=0}^{d}{\frac{D_{\infty }^{2}G_{\infty }^{DRL}\sqrt{1-{{\beta }_{2}}}}{2\alpha \ (1-{{\beta }_{1}}){{(1-\lambda )}^{2}}}}
\end{align}
The results obtained from the primary analysis of sum rate and traffic densities are shown in Fig. 5(b) in long-term settings. It shows that the proposed model performs close to the benchmark scheme and better than OMA. Additionally, we compare the proposed NOMA with dynamic power and traditional NOMA scheme with the fixed power as NOMA-fix(P). It is clearly visible that the traditional method of allocation as NOMA-fix(P) is not as efficient as the proposed allocation strategy. Fig.6(a) shows short-term performance analysis between sum rates, bandwidth, and the number of iterations. This figure illustrates the performance of DRL and SARSA according to different bandwidths, due to the fast convergence with heavy traffic as shown in Fig. 5(a) the performance of the DRL is better than SARSA. Interestingly it also shows that with the increase of the traffic density, the sum rate improves. Therefore, sum rates are proportional to the number of users/traffic density in this case. Furthermore, from Fig. 5(b) even with light traffic conditions, the sum rate of NOMA systems is higher than OMA. Lastly,  Fig. 6(b) shows the relationships among long-term users connectivity during the simulation time. Where it is clearly visible that NOMA is more efficient for user connectivity by serving more users than OMA. From this figure, we can see that the connectivity is improving as reinforcement learning agents, specifically DRL agent learning the dynamic environment. The number of served users are significantly increasing after 150 episodes of learning. As we can see the total number of served users are more than 3000 for DRL NOMA and more than 1000 for SARSA NOMA within 200 episodes.
\begin{figure}[htp!]
\centering
\subfigure[]{\label{shortb}\includegraphics[scale=.5,keepaspectratio]{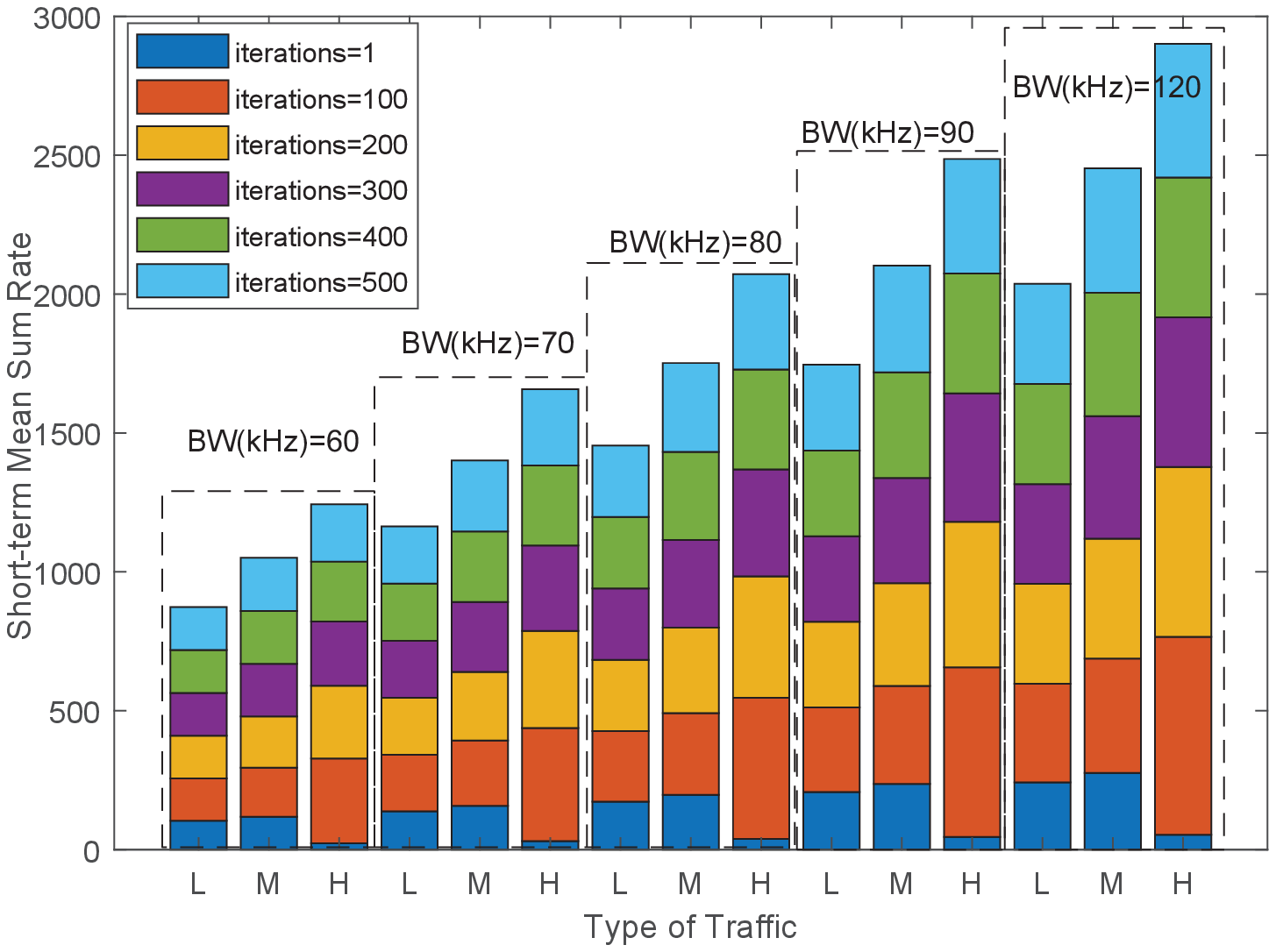}}
\subfigure[]{\label{longc}\includegraphics[scale=.5,keepaspectratio]{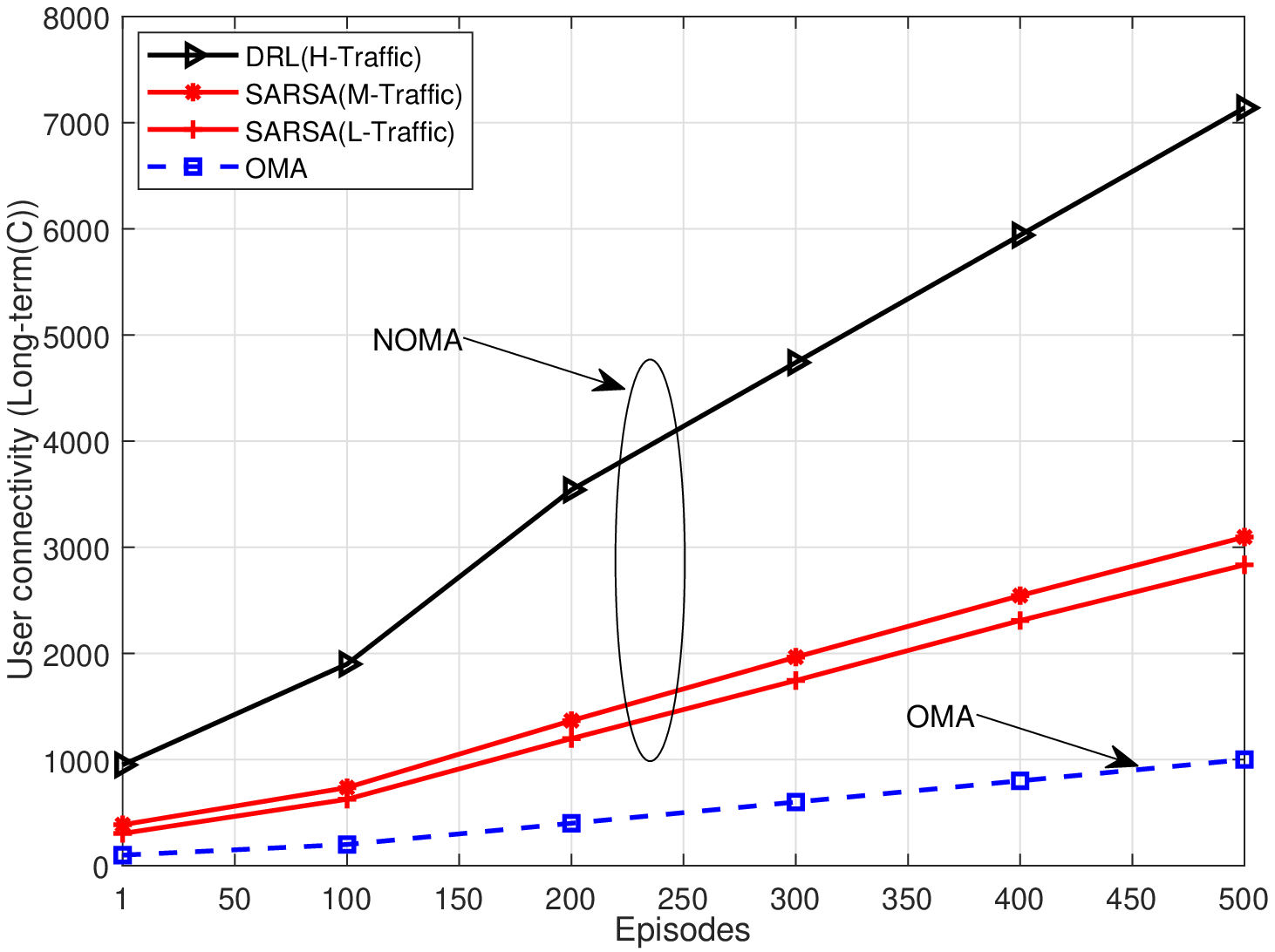}}
\caption{ Overview of the proposed framework for the sum-rate maximization problem. Sub-figure (a) is a short-term comparison between the channel bandwidth, average sum rate, and different network traffic loads for the proposed DRL and SARSA. The parameter (L) denotes light traffic, (M) denotes medium, and (H) is for heavy traffic. (b) shows a long-term comparison between time episodes and clustering parameter c, showing the sum of connected users in the long-term for the proposed SARSA, DRL and OMA scheme.}

\end{figure}
\vspace{-0.5 cm} 
\subsection{DQN Loss vs Rewards}
\begin{figure}
\centering
    \subfigure[]{\label{drlloss}\includegraphics[scale=.5,keepaspectratio]{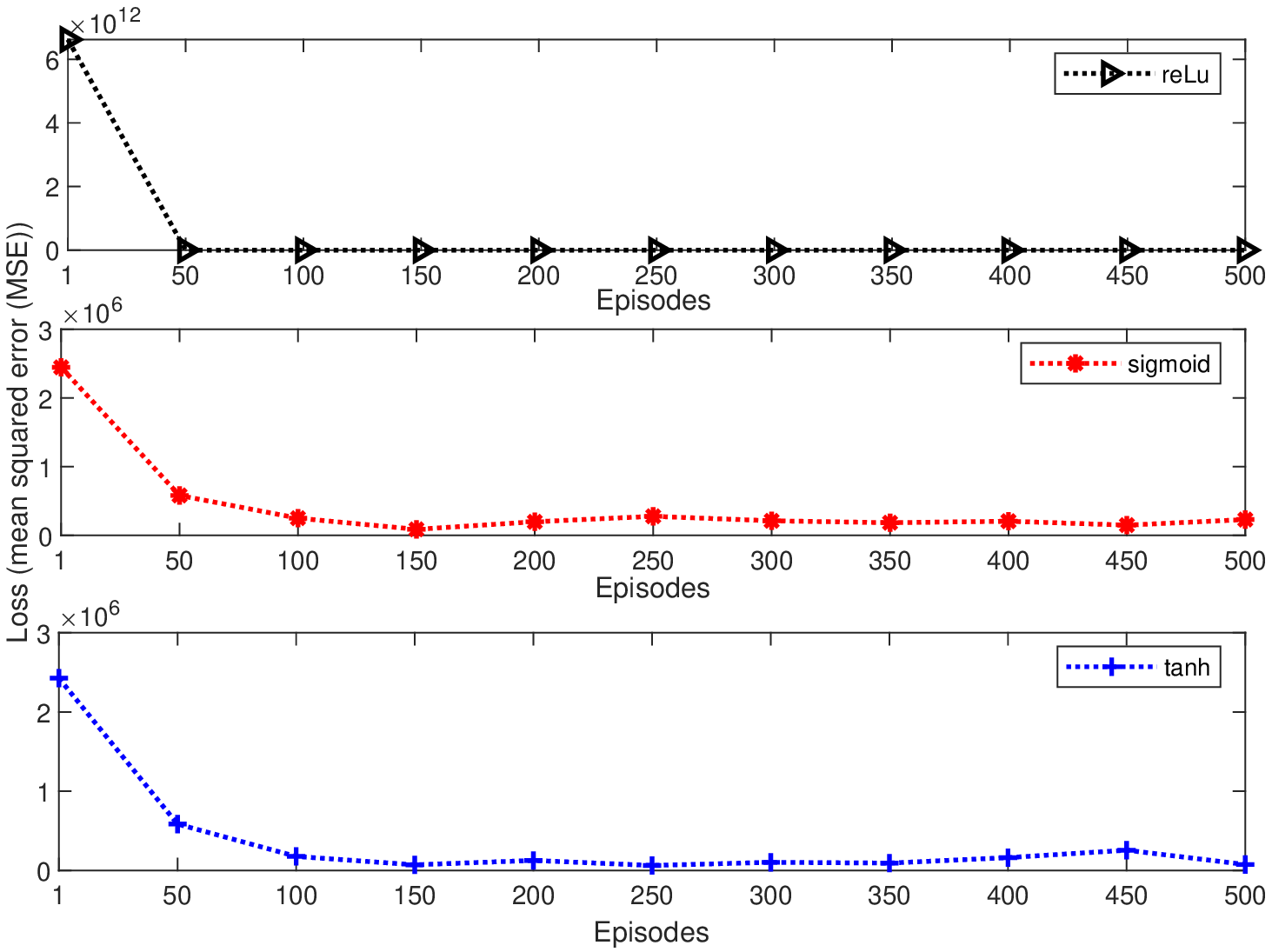}}
    \subfigure[]{\label{drlact}\includegraphics[scale=.5,keepaspectratio]{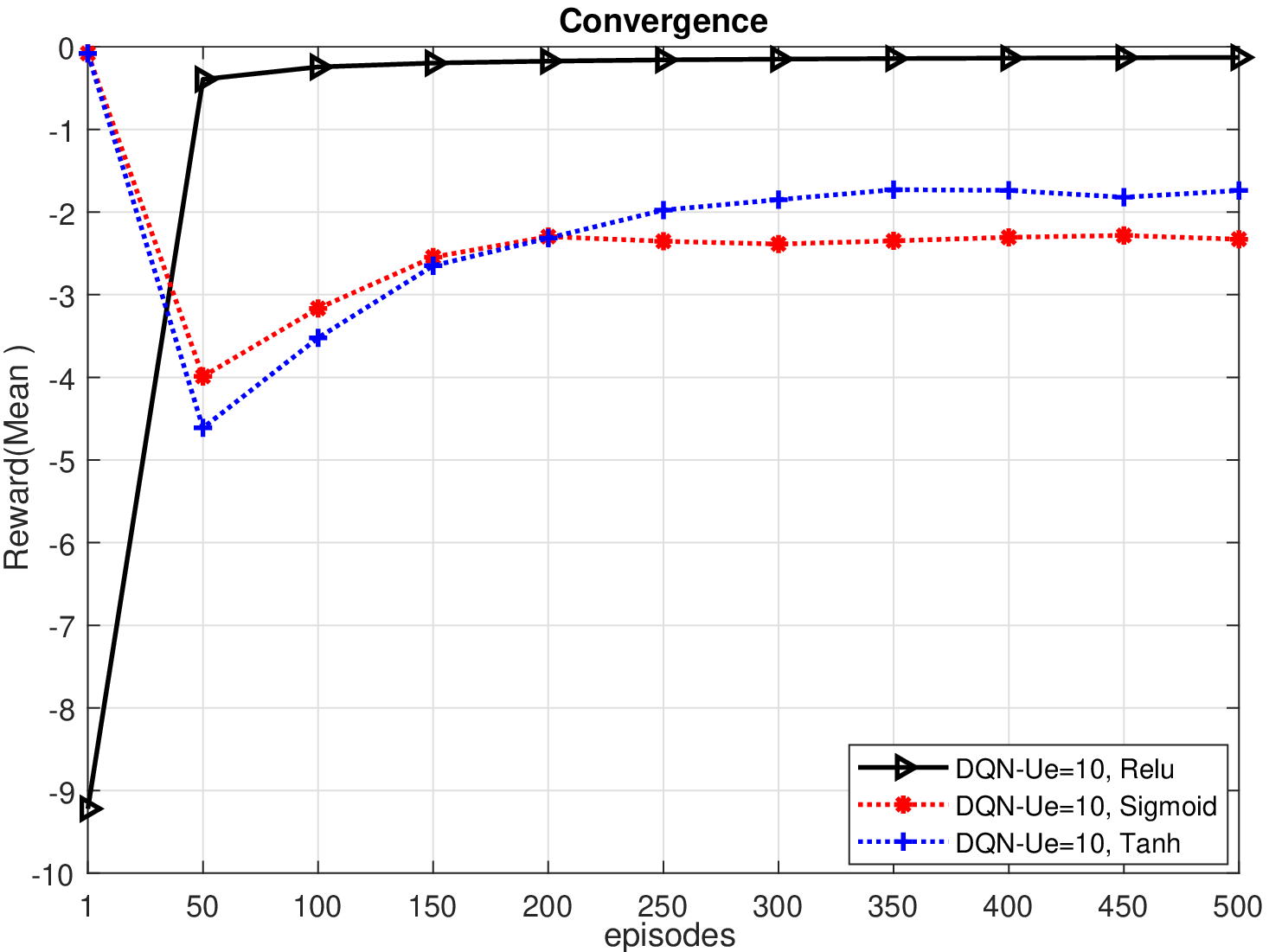}}
\caption{  Overview of the proposed framework for the sum-rate maximization problem. Sub-figure (a) DRL loss vs the number of episodes: A comparison between DRL loss and training episodes for different activation functions (ReLU, TanH, Sigmoid). Sub-figure (b) shows Rewards vs activation functions: A comparison between achieved rewards and episodes for different activation functions (ReLU, TanH, Sigmoid) of DRL algorithm.}
\vspace{-0.9 cm} 
\end{figure}

In Fig. \ref{drlloss}, the loss (MSE) for the DRL algorithm is shown, comparing three well-known activation functions (ReLU, TanH, and Sigmoid). As it can be seen that ReLU performs better than both Sigmoid and TanH activation functions. Sigmoid and TanH perform relatively better only in initial steps due to less experience of the DRL agent. Therefore, when the DRL agent gains some experience after the process of exploration and exploitation of the given environment, the outcome of the DRL algorithm is changed accordingly. The loss (y-axis) for all the activation functions is decreasing according to the number of episodes (x-axis). Furthermore, this figure also shows that the performance (loss) of the DRL algorithm is efficient when ReLU activation is used. Fig. \ref{drlact} provides the summary statistics of achieved average rewards for the three different activation functions of the DRL algorithm. From the data in Fig. \ref{drlact}, it is apparent that the DRL algorithm with ReLU outperforms Sigmoid and TanH activation functions. After combining Fig. \ref{drlloss} and Fig. \ref{drlact}, another interesting outcome is that by the improvement of the loss function, the rewards improves as well. Therefore, loss and reward are proportional to each other. Lastly, DRL with the Sigmoid activations is the second best until 200 episodes and in all the remaining cases, where the episode is greater than 200 the performance of TanH is better than Sigmoid.
\vspace{-0.5 cm} 
\subsection{Clustering Time}
\begin{figure}[htp!]
\centering
\includegraphics[scale=.5,keepaspectratio]{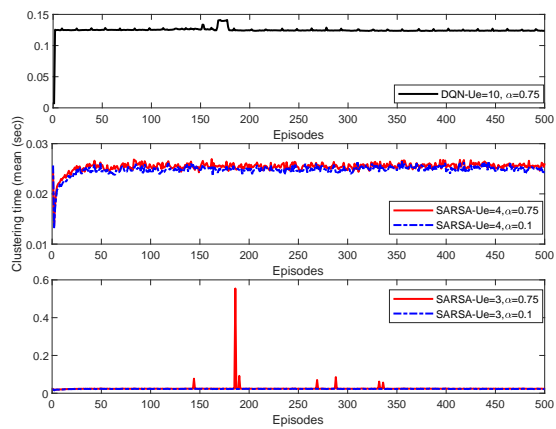}
\caption{Clustering time (mean (sec)) vs Traffic Densities for DRL and SARSA: A comparison between different traffic densities and learning rates of the proposed algorithms.}
\label{fig:8}
\vspace{-0.9 cm} 
\end{figure}

The average clustering time in second is compared for DRL and SARSA algorithms with different types of traffic and the impact of learning rates $\alpha$ in Fig. \ref{fig:8}. The learning rate is the significant hyper-parameter of RL algorithms, which shows how long the agent spends to explore and exploit the given environments. From the figure, it can be seen that there is no large effect of learning rates on clustering time (y-axis) for all the scenarios with current hyper-parameters but if it is not tuned with other hyper-parameters, the learning rate can negatively influence the learning process. Therefore, with improper tuning the learning process becomes unbalanced and the agent can be searching for the solution for an infinite amount of time. Lastly, the clustering time increases but not significant when the max load is increased from 3 to 10.
\vspace{-0.5 cm} 
\section{Conclusion}
  This paper has proposed resource allocation for IoT users in the uplink transmission of NOMA systems. Two algorithms DRL and SARSA in the present study have been designed to determine the effect of three different traffic densities on the sum rate of IoT users. In order to improve the overall sum rate under a different number of IoT users, we have formulated a multi-dimensional optimization problem using intelligent clustering based on RL algorithms with several interesting outcomes. Firstly, the simulation results of this study have indicated that the proposed technique performed close to the benchmark scheme in all the scenarios. The second major finding is that this framework provides a long-term guaranteed average rate with long-term reliability and stability. Thirdly, it has proved that DRL is efficient for complex scenarios. Additionally, we have proved that the sparse activations improve the performance of the DNNs when compared to the traditional mechanisms. Therefore, DRL with sparse activations is suitable for heavy traffic and SARSA is suitable for light traffic conditions. Furthermore, in general, both the algorithms (DRL and SARSA) have obtained better sum rates than OMA systems.  Lastly, further research will explore performance improvements under the different scale of the networks.
\begin{appendices}
\section{Proof of Problem \eqref{h}}
\numberwithin{equation}{section}
\setcounter{equation}{0}
With the aid of the theory of computation complexity, we are able to use the following two steps to prove that the problem (\ref{h}) is an NP-hard problem. Step 1: the association problem for every subset of  $\mathbf{\Phi_u^{i,j}}$ is NP-complete. Step 2: this step is to prove the relationship of $u_k^{i,j}$ and the problem in \eqref{h2} is similar to our objective function. The problem (\ref{h}) in this paper is NP-hard, following proof can be divided into two cases, namely $N_u=1$ (static clustering/association) and $N_u \geq 1$ (dynamic clustering/associations).
\begin{enumerate}
\item For the case $N_u=1$ (static clustering/association), the problem (\ref{h}) is similar to the conventional OMA systems so that the resource management problem can be expressed as follows:
\begin{align}
\label{h2} \underset{\mathbf{C},\mathbf{P}}{\max} \ \ & \mathbb{E}[ R_{sum}(t)],\\
 \mathrm{s.t:}\ \ \ &{2}\le c^{i,j}_{k}(t), \forall i ,k=1, \\
& {\sum_{k=1}^{N_u^{i,j}}{c^{i,j}_{k}(t) p^{i,j}_{k}(t) \le P_b ,\forall i,\forall j}}.
\end{align}
The above-mentioned problem has been proved to be NP-hard in \cite{naparstek2018deep} for OMA systems.
\item For the case $N_u >1$ (dynamic clustering/associations), even with known power allocations we show that the problem (\ref{h}) is NP-hard since the optimal power selection for multiple users is NP-hard. Additionally, it is known that 3D associations are NP-hard problems~\cite{cui2018optimal}. Under the condition that $N_u > 1$, for any $u_k^{i,j}$, there are more than one combinations in the set $\Phi_u^{i,j}$ even for the 3D association problem in OMA systems. Moreover, the combinations in NOMA is larger than those in OMA.
\end{enumerate}
As a result, the decision problem of the constructed instance is NP-complete and the main instance is NP-hard.
\end{appendices}

\bibliographystyle{IEEEtran}
\bibliography{mybib}

\begin{thebibliography}{10}
\providecommand{\url}[1]{#1}
\csname url@samestyle\endcsname
\providecommand{\newblock}{\relax}
\providecommand{\bibinfo}[2]{#2}
\providecommand{\BIBentrySTDinterwordspacing}{\spaceskip=0pt\relax}
\providecommand{\BIBentryALTinterwordstretchfactor}{4}
\providecommand{\BIBentryALTinterwordspacing}{\spaceskip=\fontdimen2\font plus
\BIBentryALTinterwordstretchfactor\fontdimen3\font minus
  \fontdimen4\font\relax}
\providecommand{\BIBforeignlanguage}[2]{{%
\expandafter\ifx\csname l@#1\endcsname\relax
\typeout{** WARNING: IEEEtran.bst: No hyphenation pattern has been}%
\typeout{** loaded for the language `#1'. Using the pattern for}%
\typeout{** the default language instead.}%
\else
\language=\csname l@#1\endcsname
\fi
#2}}
\providecommand{\BIBdecl}{\relax}
\BIBdecl

\bibitem{123456789}
W.~Ahsan, W.~Yi, Z.~Qin, Y.~Liu, and A.~Nallanathan, ``Reinforcement learning
  for user clustering in {NOMA}-enabled uplink {IoT},'' in \emph{IEEE Proc. of
  International Commun. Conf. (ICC Wkshps)}, Jun. 2020.

\bibitem{zhai2019delay}
D.~Zhai, R.~Zhang, L.~Cai, and F.~R. Yu, ``Delay minimization for massive
  internet of things with non-orthogonal multiple access,'' \emph{{IEEE} J.
  Sel. Areas Commun.}, vol.~13, no.~3, pp. 553--566, 2019.

\bibitem{islam2017power}
S.~R. Islam, N.~Avazov, O.~A. Dobre, and K.-S. Kwak, ``Power-domain
  non-orthogonal multiple access {(NOMA)} in {5G} systems: Potentials and
  challenges,'' \emph{{IEEE} Commun. Surveys Tuts.}, vol.~19, no.~2, pp.
  721--742, 2017.

\bibitem{sharma2019towards}
S.~K. {Sharma} and X.~{Wang}, ``Towards massive machine type communications in
  ultra-dense cellular iot networks: Current issues and machine
  learning-assisted solutions,'' \emph{{IEEE} Commun. Surveys Tuts.}, pp. 1--1,
  2019.

\bibitem{wan2018non}
D.~Wan, M.~Wen, F.~Ji, H.~Yu, and F.~Chen, ``Non-orthogonal multiple access for
  cooperative communications: Challenges, opportunities, and trends,''
  \emph{{IEEE} Wireless Commun.}, vol.~25, no.~2, pp. 109--117, 2018.

\bibitem{guo2020weighted}
H.~Guo, Y.-C. Liang, J.~Chen, and E.~G. Larsson, ``Weighted sum-rate
  maximization for reconfigurable intelligent surface aided wireless
  networks,'' \emph{{IEEE} Trans. Wireless Commun.}, vol.~19, no.~5, pp.
  3064--3076, 2020.

\bibitem{zeng2020sum}
M.~Zeng, X.~Li, G.~Li, W.~Hao, and O.~Dobre, ``Sum rate maximization for
  irs-assisted uplink noma,'' \emph{arXiv preprint arXiv:2004.10791}, 2020.

\bibitem{tse2005fundamentals}
D.~Tse and P.~Viswanath, \emph{Fundamentals of wireless communication}.\hskip
  1em plus 0.5em minus 0.4em\relax Cambridge university press, 2005.

\bibitem{ding2017survey}
Z.~Ding, X.~Lei, G.~K. Karagiannidis, R.~Schober, J.~Yuan, and V.~K. Bhargava,
  ``A survey on non-orthogonal multiple access for {5G} networks: Research
  challenges and future trends,'' \emph{{IEEE} J. Sel. Areas Commun.}, vol.~35,
  no.~10, pp. 2181--2195, 2017.

\bibitem{shao2018dynamic}
X.~Shao, C.~Yang, D.~Chen, N.~Zhao, and F.~R. Yu, ``Dynamic {IoT} device
  clustering and energy management with hybrid {NOMA} systems,'' \emph{{IEEE}
  Trans. Ind. Informat.}, vol.~14, no.~10, pp. 4622--4630, 2018.

\bibitem{ali2016dynamic}
M.~S. Ali, H.~Tabassum, and E.~Hossain, ``Dynamic user clustering and power
  allocation for uplink and downlink non-orthogonal multiple access (noma)
  systems,'' \emph{IEEE access}, vol.~4, pp. 6325--6343, 2016.

\bibitem{miuccio2020joint}
L.~Miuccio, D.~Panno, and S.~Riolo, ``Joint control of random access and
  dynamic uplink resource dimensioning for massive mtc in 5g nr based on
  scma,'' \emph{{IEEE} Internet Things J.}, 2020.

\bibitem{mostafa2019connection}
A.~E. Mostafa, Y.~Zhou, and V.~W. Wong, ``Connection density maximization of
  narrowband iot systems with {NOMA},'' \emph{{IEEE} Trans. Wireless Commun.},
  vol.~18, no.~10, pp. 4708--4722, 2019.

\bibitem{8519960}
D.~{Wang}, D.~{Chen}, B.~{Song}, N.~{Guizani}, X.~{Yu}, and X.~{Du}, ``From
  {IoT} to {5G} {I-IoT}: The next generation {IoT-Based} intelligent algorithms
  and {5G} technologies,'' \emph{{IEEE} Commun. Mag.}, vol.~56, no.~10, pp.
  114--120, OCTOBER 2018.

\bibitem{hussain2019machine}
F.~Hussain, S.~A. Hassan, R.~Hussain, and E.~Hossain, ``Machine learning for
  resource management in cellular and {IoT} networks: Potentials, current
  solutions, and open challenges,'' \emph{arXiv preprint arXiv:1907.08965},
  2019.

\bibitem{zhang2016uplink}
N.~Zhang, J.~Wang, G.~Kang, and Y.~Liu, ``Uplink nonorthogonal multiple access
  in {5G} systems,'' \emph{{IEEE} Commun. Lett.}, vol.~20, no.~3, pp. 458--461,
  2016.

\bibitem{ding2014performance}
Z.~Ding, Z.~Yang, P.~Fan, and H.~V. Poor, ``On the performance of
  non-orthogonal multiple access in {5G} systems with randomly deployed
  users,'' \emph{arXiv preprint arXiv:1406.1516}, 2014.

\bibitem{hanif2016minorization}
M.~F. Hanif, Z.~Ding, T.~Ratnarajah, and G.~K. Karagiannidis, ``A
  minorization-maximization method for optimizing sum rate in the downlink of
  non-orthogonal multiple access systems.'' \emph{IEEE Trans. Signal Process.},
  vol.~64, no.~1, pp. 76--88, 2016.

\bibitem{zhang2018energy}
H.~Zhang, B.~Wang, C.~Jiang, K.~Long, A.~Nallanathan, V.~C. Leung, and H.~V.
  Poor, ``Energy efficient dynamic resource optimization in {NOMA} system,''
  \emph{{IEEE} Trans. Wireless Commun.}, vol.~17, no.~9, pp. 5671--5683, 2018.

\bibitem{yang2016general}
Z.~Yang, Z.~Ding, P.~Fan, and N.~Al-Dhahir, ``A general power allocation scheme
  to guarantee quality of service in downlink and uplink {NOMA} systems,''
  \emph{{IEEE} Trans. Wireless Commun.}, vol.~15, no.~11, pp. 7244--7257, 2016.

\bibitem{zhai2018energy}
D.~Zhai, R.~Zhang, L.~Cai, B.~Li, and Y.~Jiang, ``Energy-efficient user
  scheduling and power allocation for {NOMA}-based wireless networks with
  massive {IoT} devices,'' \emph{{IEEE} Internet Things J.}, vol.~5, no.~3, pp.
  1857--1868, 2018.

\bibitem{liu2017enhancing}
Y.~Liu, Z.~Qin, M.~Elkashlan, Y.~Gao, and L.~Hanzo, ``Enhancing the physical
  layer security of non-orthogonal multiple access in large-scale networks.''
  \emph{{IEEE} Trans.Wireless Commun.}, vol.~16, no.~3, pp. 1656--1672, 2017.

\bibitem{8635489}
W.~{Yi}, Y.~{Liu}, A.~{Nallanathan}, and M.~{Elkashlan}, ``Clustered
  millimeter-wave networks with non-orthogonal multiple access,'' \emph{{IEEE}
  Trans. Commun.}, vol.~67, no.~6, pp. 4350--4364, Jun. 2019.

\bibitem{ali2018coordinated}
M.~S. Ali, E.~Hossain, and D.~I. Kim, ``Coordinated multipoint transmission in
  downlink multi-cell {NOMA} systems: Models and spectral efficiency
  performance,'' \emph{{IEEE} Wireless Commun.}, vol.~25, no.~2, pp. 24--31,
  2018.

\bibitem{qian2018optimal}
L.~P. {Qian}, A.~{Feng}, Y.~{Huang}, Y.~{Wu}, B.~{Ji}, and Z.~{Shi}, ``Optimal
  {SIC} ordering and computation resource allocation in {MEC}-aware {NOMA}
  {NB-IoT} networks,'' \emph{{IEEE} Internet Things J.}, vol.~6, no.~2, pp.
  2806--2816, April 2019.

\bibitem{shahab2019grant}
M.~B. Shahab, R.~Abbas, M.~Shirvanimoghaddam, and S.~J. Johnson, ``Grant-free
  non-orthogonal multiple access for {IoT}: A survey,'' \emph{arXiv preprint
  arXiv:1910.06529}, 2019.

\bibitem{dai2018survey}
L.~{Dai}, B.~{Wang}, Z.~{Ding}, Z.~{Wang}, S.~{Chen}, and L.~{Hanzo}, ``A
  survey of non-orthogonal multiple access for {5G},'' \emph{{IEEE} Commun.
  Surveys Tuts.}, vol.~20, no.~3, pp. 2294--2323, thirdquarter 2018.

\bibitem{gui2018deep}
G.~Gui, H.~Huang, Y.~Song, and H.~Sari, ``Deep learning for an effective
  nonorthogonal multiple access scheme,'' \emph{{IEEE} Trans. Veh. Technol.},
  vol.~67, no.~9, pp. 8440--8450, 2018.

\bibitem{hochreiter1997long}
S.~Hochreiter and J.~Schmidhuber, ``Long short-term memory,'' \emph{Neural
  computation}, vol.~9, no.~8, pp. 1735--1780, 1997.

\bibitem{xu2018outage}
Y.~Xu, D.~Cai, F.~Fang, Z.~Ding, C.~Shen, and G.~Zhu, ``Outage analysis and
  power allocation for {HARQ-CC} enabled {NOMA} downlink transmission,'' in
  \emph{{IEEE} Glob. Commun. Conf. {(GLOBECOM)}}, Dec 2018, pp. 1--6.

\bibitem{jiang2017machine}
C.~Jiang, H.~Zhang, Y.~Ren, Z.~Han, K.-C. Chen, and L.~Hanzo, ``Machine
  learning paradigms for next-generation wireless networks,'' \emph{{IEEE}
  Wireless Commun.}, vol.~24, no.~2, pp. 98--105, 2017.

\bibitem{bi2015wireless}
S.~Bi, R.~Zhang, Z.~Ding, and S.~Cui, ``Wireless communications in the era of
  big data,'' \emph{arXiv preprint arXiv:1508.06369}, 2015.

\bibitem{arafat2019localization}
M.~Y. {Arafat} and S.~{Moh}, ``Localization and clustering based on swarm
  intelligence in {UAV} networks for emergency communications,'' \emph{{IEEE}
  Internet Things J.}, vol.~6, no.~5, pp. 8958--8976, Oct 2019.

\bibitem{cui2018unsupervised}
J.~Cui, Z.~Ding, P.~Fan, and N.~Al-Dhahir, ``Unsupervised machine
  learning-based user clustering in millimeter-wave-{NOMA} systems,''
  \emph{{IEEE} Trans. Wireless Commun.}, vol.~17, no.~11, pp. 7425--7440, 2018.

\bibitem{liu2019machine}
Y.~Liu, S.~Bi, Z.~Shi, and L.~Hanzo, ``When machine learning meets big data: A
  wireless communication perspective,'' \emph{arXiv preprint arXiv:1901.08329},
  2019.

\bibitem{li2020multi}
F.~Li, D.~Yu, H.~Yang, J.~Yu, H.~Karl, and X.~Cheng, ``Multi-armed-bandit-based
  spectrum scheduling algorithms in wireless networks: A survey,'' \emph{{IEEE}
  Wireless Commun.}, vol.~27, no.~1, pp. 24--30, 2020.

\bibitem{de2018comparing}
T.~B. de~Oliveira, A.~L. Bazzan, B.~C. da~Silva, and R.~Grunitzki, ``Comparing
  multi-armed bandit algorithms and q-learning for multiagent action selection:
  a case study in route choice,'' in \emph{2018 International Joint Conference
  on Neural Networks {(IJCNN)}}.\hskip 1em plus 0.5em minus 0.4em\relax IEEE,
  2018, pp. 1--8.

\bibitem{silver2017mastering}
D.~Silver, J.~Schrittwieser, K.~Simonyan, I.~Antonoglou, A.~Huang, A.~Guez,
  T.~Hubert, L.~Baker, M.~Lai, A.~Bolton \emph{et~al.}, ``Mastering the game of
  go without human knowledge,'' \emph{Nature}, vol. 550, no. 7676, p. 354,
  2017.

\bibitem{rummery1994line}
G.~A. Rummery and M.~Niranjan, \emph{On-line Q-learning using connectionist
  systems}.\hskip 1em plus 0.5em minus 0.4em\relax University of Cambridge,
  Department of Engineering Cambridge, England, 1994, vol.~37.

\bibitem{lillicrap2015continuous}
T.~P. Lillicrap, J.~J. Hunt, A.~Pritzel, N.~Heess, T.~Erez, Y.~Tassa,
  D.~Silver, and D.~Wierstra, ``Continuous control with deep reinforcement
  learning,'' \emph{arXiv preprint arXiv:1509.02971}, 2015.

\bibitem{xiao2017reinforcement}
L.~Xiao, Y.~Li, C.~Dai, H.~Dai, and H.~V. Poor, ``Reinforcement learning-based
  {NOMA} power allocation in the presence of smart jamming,'' \emph{{IEEE}
  Trans. Veh. Technol.}, vol.~67, no.~4, pp. 3377--3389, 2017.

\bibitem{liu2019uav}
Y.~Liu, Z.~Qin, Y.~Cai, Y.~Gao, G.~Y. Li, and A.~Nallanathan, ``{UAV}
  communications based on non-orthogonal multiple access,'' \emph{{IEEE}
  Wireless Commun.}, vol.~26, no.~1, pp. 52--57, 2019.

\bibitem{yang2019reinforcement}
H.~Yang, P.~Du, W.-D. Zhong, C.~Chen, A.~Alphones, and S.~Zhang,
  ``Reinforcement learning based intelligent resource allocation for integrated
  {VLCP} systems,'' \emph{{IEEE} Wireless Commun. Lett.}, vol.~8, no.~4, pp.
  1204--1207, 2019.

\bibitem{cui2017optimal}
J.~Cui, Y.~Liu, Z.~Ding, P.~Fan, and A.~Nallanathan, ``Optimal user scheduling
  and power allocation for millimeter wave {NOMA} systems,'' \emph{{IEEE}
  Trans. Wireless Commun.}, vol.~17, no.~3, pp. 1502--1517, 2017.

\bibitem{kiani2018edge}
A.~Kiani and N.~Ansari, ``Edge computing aware {NOMA} for {5G} networks,''
  \emph{{IEEE} Internet Things J.}, vol.~5, no.~2, pp. 1299--1306, 2018.

\bibitem{liu2016cooperative}
Y.~Liu, Z.~Ding, M.~Elkashlan, and H.~V. Poor, ``Cooperative non-orthogonal
  multiple access with simultaneous wireless information and power transfer,''
  \emph{{IEEE} J. Sel. Areas Commun.}, vol.~34, no.~4, pp. 938--953, 2016.

\bibitem{803503}
{Qinqing Zhang} and S.~A. {Kassam}, ``Finite-state {Markov} model for rayleigh
  fading channels,'' \emph{{IEEE} Trans. Commun.}, vol.~47, no.~11, pp.
  1688--1692, Nov 1999.

\bibitem{8680645}
F.~{Fang}, Z.~{Ding}, W.~{Liang}, and H.~{Zhang}, ``Optimal energy efficient
  power allocation with user fairness for uplink {MC-NOMA} systems,''
  \emph{{IEEE} Wireless Commun. Lett.}, pp. 1--1, 2019.

\bibitem{8626185}
X.~{Liu}, Z.~{Qin}, Y.~{Gao}, and J.~A. {McCann}, ``Resource allocation in
  wireless powered {IoT} networks,'' \emph{{IEEE} Internet Things J.}, vol.~6,
  no.~3, pp. 4935--4945, June 2019.

\bibitem{cui2018optimal}
J.~Cui, Y.~Liu, Z.~Ding, P.~Fan, and A.~Nallanathan, ``Optimal user scheduling
  and power allocation for millimeter wave {NOMA} systems,'' \emph{{IEEE}
  Trans. Wireless Commun.}, vol.~17, no.~3, pp. 1502--1517, 2018.

\bibitem{8807386}
J.~{Cui}, Y.~{Liu}, and A.~{Nallanathan}, ``Multi-agent reinforcement
  learning-based resource allocation for {UAV} networks,'' \emph{{IEEE} Trans.
  Wireless Commun.}, vol.~19, no.~2, pp. 729--743, Feb 2020.

\bibitem{watkins1992q}
C.~J. Watkins and P.~Dayan, ``{Q}-learning,'' \emph{Machine learning}, vol.~8,
  no. 3-4, pp. 279--292, 1992.

\bibitem{melo2001convergence}
F.~S. Melo, ``Convergence of {Q}-learning: A simple proof,'' \emph{Institute Of
  Systems and Robotics, Tech. Rep}, pp. 1--4, 2001.

\bibitem{zhang1999finite}
Q.~Zhang and S.~A. Kassam, ``Finite-state {Markov} model for rayleigh fading
  channels,'' \emph{{IEEE} Trans. Commun.}, vol.~47, no.~11, pp. 1688--1692,
  1999.

\bibitem{kingma2014adam}
D.~P. Kingma and J.~Ba, ``Adam: A method for stochastic optimization,''
  \emph{arXiv preprint arXiv:1412.6980}, 2014.

\bibitem{naparstek2018deep}
O.~Naparstek and K.~Cohen, ``Deep multi-user reinforcement learning for
  distributed dynamic spectrum access,'' \emph{{IEEE} Trans. Wireless Commun.},
  vol.~18, no.~1, pp. 310--323, 2018.

\end{thebibliography}

\end{document}